\documentclass[fleqn,usenatbib]{mnras}

\usepackage{newtxtext,newtxmath}

\usepackage[T1]{fontenc}

\DeclareRobustCommand{\VAN}[3]{#2}
\let\VANthebibliography\thebibliography
\def\thebibliography{\DeclareRobustCommand{\VAN}[3]{##3}\VANthebibliography}

\usepackage{graphicx}	
\usepackage{amsmath}	

\title[MHD Meridional Circulation]{Meridional Circulation driven by Planetary Spiral Wakes in Radiative and Magnetized Protoplanetary Discs}

\author[Cilibrasi et al.]{
Marco Cilibrasi,$^{1,2}$\thanks{E-mail: marco.cilibrasi@uzh.ch}
Mario Flock,$^{3}$
Judit Szul{\'a}gyi$^{2}$
\\
$^{1}$Institute for Computational Science, University of Z\"urich, Winterthurerstrasse 190, CH-8057 Z\"urich, Switzerland\\
$^{2}$ETH Z\"urich, Department of Physics, Wolfgang-Pauli-Strasse 27, CH-8093, Z\"urich, Switzerland\\
$^{3}$Max-Planck-Institut f\"ur Astronomie, K\"onigstuhl 17, D-69117 Heidelberg, Germany
}

\date{Accepted XXX. Received YYY; in original form ZZZ}

\pubyear{2022}

\begin{document}
\label{firstpage}
\pagerange{\pageref{firstpage}--\pageref{lastpage}}
\maketitle

\begin{abstract}
We study a Jupiter-mass planet formation for the first time in radiative magneto-hydrodynamics (MHD) simulations and compare it with pure hydrodynamical simulations, as well as to different isothermal configurations. We found that the meridional circulation is the same in every setup. The planetary spiral wakes drive a vertical stirring inside the protoplanetary disc and the encounter with these shock fronts also helps in delivering gas vertically onto the Hill-sphere. The accretion dynamics are unchanged: the planet accretes vertically, and there is outflow in the midplane regions inside the Hill-sphere. We determined the effective $\alpha$-viscosity generated in the disc by the various angular momentum loss mechanisms, which showed that magnetic fields produce high turbulence in the ideal MHD limit, that grows from $\alpha \sim 10^{-2.5}$ up to $\sim 10^{-1.5}$ after the planet spirals develop. In the HD simulations, the planetary spirals contribute to $\alpha \sim 10^{-3}$, making this a very important angular momentum transport mechanism. Due to the various $\alpha$ values in the different setups, the gap opening is different in each case. In the radiative MHD setups, the high turbulent viscosity prevents gap opening, leading to a higher Hill mass, and no clear dust trapping regions. While the Hill accretion rate is $10^{-6} \rm{M_{Jup}/yr}$ in all setups, the accretion variability is orders of magnitude higher in radiative runs than in isothermal ones. Finally, with higher-resolution runs, the magneto-rotational instability started to be resolved, changing the effective viscosity and increasing the heating in the disc.
\end{abstract}




\section{Introduction}\label{sec:intro}

Over the last decades, a lot of work has been done on the structure and evolution of protoplanetary discs (PPDs), and their interaction with a forming planet, both analytically and numerically \citep{Goldreich80, Lin86, Kley12}. In particular, it is now well known that the disc exerts a torque on a forming planet, causing, first of all, an exchange of angular momentum that reflects into a change of the planetary semi-major axis, i.e. a migration, usually referred to as Type I \citep{Ward97, Tanaka02, Paardekooper06, Jimenez17}. If the mass of the planet is large enough, one or multiple annular gaps can open around the orbit of the planet, i.e. circular regions where the density drops compared to the surrounding disc. As soon as the gap is opened, the angular momentum exchange stops, and migration is now dominated by the dynamics of the disc itself, i.e. the planet enters a regime of Type II migration \citep{Syer95, Nelson00, Armitage05}.

In more detail, a gap is opened because the structure of the protoplanetary disc adjusts to the angular momentum exchange, with the result that gas is pushed away from the planetary orbit by the torques. At the same time, viscous forces tend to fill the gap back in, hence the gap opening process is controlled by the counteractive effects of these torques, together with co-orbital torques due to the non-axisymmetric structure of the wakes around the planet \citep{Goldreich80, Lin80, Papaloizou84}. In recent years, many other torque components have been found, for example, the so-called \textit{heating torque} that would prevent or slow down inward migration of a planet \citep{Benitez15} due to the increased temperature caused by accretion heating in the  protoplanet immediate vicinity. Complex thermodynamics could lead also to more peculiar phenomena, such as the formation of a cold finger of gas that again counteracts migration \citep{Lega14, Masset17}. The migration process depends then on planetary properties, such as the mass, but also on the disc's properties, such as the temperature and the viscosity, with the result that gaps are more likely opened in low-viscosity and low-temperature discs, or in the case of larger planets \citep{Crida06, Armitage10}. 

Even when a planetary gap is fully opened, the gas dynamics around the planet itself would not completely decouple from the protoplanetary disc. In fact, some gas would flow through the gap and reach the circumplanetary region, as well as the inner circumstellar disc. As the flow enters the gap region, the gas parcels encountering the planetary spiral wake (which is a shock-front) will lose angular momentum when passing through the shock and hence will spiral down to the circumplanetary region. Whether a circumplanetary envelope or a  circumplanetary disc (CPD) forms \citep{Kley99,Lubow99, Szulagyi14, Szulagyi16}, depends on the planetary mass as well as the temperature and opacity in the planet's immediate vicinity. The higher the planetary mass and the cooler is the Hill-sphere, the more likely there is a disc around the planet, instead of an envelope \citep{Szulagyi16}.  Even if circumplanetary discs may seem just descaled versions of protoplanetary discs, there are several differences between them. So far their characteristics have been investigated mainly with semi-analytical models \citep{Canup02, Alibert05, Canup06, Ward10} and hydrodynamical simulations \citep{Ayliffe09a, Ayliffe09b, Gressel13, Shabram13, Zhu14, Szulagyi14, Szulagyi16,Szulagyi17a, Szulagyi19}. In particular, the most important difference is that evolved PPDs are independent structures that, at least in their late stages, can be assumed to be isolated from their original star-forming cloud, while CPDs are continuously fed by a flow of material (gas and dust) from the PPDs, coming from high latitudes and accreting vertically \citep{Tanigawa12, Szulagyi14}. This is part of the meridional circulation, driven by the planet's spiral wakes, that is not only bringing in material to the planet's vicinity but also mixing up the gas and dust in the circumstellar disc \citep{Szulagyi14, Teague19, Szulagyi19, Szulagyi21}. The vertical gas flows towards the CPD and the planet, but the majority of the gas, which could not be accreted right away by the planet, will spiral outward from the CPD \citep{Szulagyi14}. 

Differences have been found between isothermal and radiative simulations (i.e. including heating and cooling effects). The heating in the forming planet vicinity will influence the accretion rates and the torques \citep{Ayliffe09a, Szulagyi16, Cimerman17, Szulagyi21}, and the vertical temperature profile has been found to determine the circulation pattern of the gas \citep{Philippov17}. Even opacity is of great importance when dealing with accretion rates, with lower opacities being responsible of higher accretion rates \citep{Papaloizou05, Ayliffe09b}. Furthermore, protoplanetary discs have been found to present a wide range of viscosity, quantified through the $\alpha$ parameter defined by \cite{Shakura73}, namely between $10^{-4}$ and $0.1$ \citep{Rafikov17}. Finally, another important dependence of the accretion rate is on the mass of the protoplanetary disc. In fact, planetary accretion rates are found to depend linearly or nearly linearly on the PPD mass \citep{Ayliffe09a}. The meridional circulation involves also dust and solid particles. In fact, a gap is opened also in the dust profile in the case of small and coupled particles \citep{Dipierro17, Binkert21}, that can also be trapped more easily around the planet \citep{Szulagyi22}.

Magneto-HydroDynamics (MHD) solvers have been widely used to study the structure and the dynamics of protoplanetary discs \citep{Lyra08, Lesur22}. For example, a series of papers \citep{Papaloizou03, Nelson03} highlighted that MHD turbulence sources, in particular, magnetorotational instability (MRI) \citep{Balbus98}, would cause a Keplerian disc to present an averaged stress parameter $\alpha\sim5\times10^{-3}$ and a magnetic $\beta$ parameter (i.e. the ratio between thermal and magnetic pressure) around 100. On the other hand, in an isothermal scenario, a planet would always open a clear gap, with no evident net mass flux towards the planet. In fact, the gap has been found to be deeper and wider in the magnetic case, as if the disc possessed a smaller effective viscosity, because of the efficient magnetic field transport through the gap. This was confirmed by \cite{Zhu13}, that performed HD and MHD simulations in an isothermal case and with a low-mas planet forming in the disc. Furthermore, \cite{Uribe11} found that in magnetized discs dominated by the Magneto-Rotational Instability (MRI), in the case of low-mass planets, the migration is dominated by random fluctuations of the torque, while for Jupiter-like planets, migration rates are similar to the ones in viscous discs. For intermediate-mass planets though, an unexpected outward migration often takes place. Other global MHD simulations of protoplanetary discs were performed by \cite{Flock13, Flock17}, which found that the magnetorotational turbulence would cause the $\alpha$ value to be very high in the inner disc, up to $0.1$, that then decreases down to $10^{-3}$ away from the star, where the disc has the dead zone. 

MHD was first implemented in the study of Circumplanetary Discs by \cite{Gressel13}, which includes Ohmic resistivity in their model, resulting in turbulent surface layers in the CPD, with a dead zone developing in the mid-plane. \cite{Fujii14} also underlined that MRI should not play a major role in CPDs, given their low ionization rate
More recently, \cite{Aoyama23} performed MHD simulations of giant planets embedded in the outer protoplanetary disc, where the Type II migration regime is fully reached. On one hand, they implemented a simple local isothermal model, on the other hand, they solved for non-ideal MHD effects, such as Ohmic resistivity and the dominant ambipolar diffusion. By resolving MRI and magnetic winds, they found that the gap is much deeper in the MHD case, while the meridional flow is weakened.

In this work we present for the first time 3D radiative MHD simulations of planet formation, focusing on the meridional circulation, the Hill-sphere accretion rates, the planetary gap opening, and the effective viscosities for the various angular momentum loss mechanisms. The radiative transfer equations were solved via the flux-limited diffusion approximation following the implementation of \cite{Kley89, Commercon11, Flock13, Szulagyi16, Flock17} in the PLUTO code \citep{Mignone07}. A global PPD has been simulated in with different resolutions, and various initial magnetic field strengths. We will compare the MHD and the HD dynamics, accretion rates, and the sources of viscosity. In all simulations, the planetary mass was one Jupiter mass.

\section{Methods and Setup}\label{sec:methods}

In this section, we present the numerical methods and the setup that have been used to perform radiative MHD simulations of protoplanetary discs with the code PLUTO \citep{Mignone07, Mignone12a, Mignone12b, Mignone14, Kolb13}.

\subsection{Physical Model}\label{sec:physical_model}

In this work, we solve the ideal MHD equations including radiative transfer in a spherical coordinate system ($r$, $\theta$, $\phi$), i.e.

\begin{equation}\label{eq:global}
    \begin{cases}
        \frac{\partial \rho}{\partial t}  + \nabla \cdot \left(\rho \mathbf{v} \right) = 0\\[3ex]
        \frac{\partial}{\partial t}\left(\rho \mathbf{v} \right) + \nabla \cdot \left(\rho \mathbf{v} \mathbf{v} - \mathbf{B} \mathbf{B} + P_t \mathbb{I}\right) = -\rho \nabla \Phi\\[3ex]
        \frac{\partial E_t}{\partial t}  + \nabla \cdot \left[(E_t+P_t) \mathbf{v} - (\mathbf{v}\times\mathbf{B})\times\mathbf{B}\right] = \\
        -\rho\mathbf{v}\times \nabla\Phi - \kappa_P \rho c (a_RT^4-E_R)\\[3ex]
        \frac{\partial E_R}{\partial t}  -\nabla\frac{c\lambda}{k_R\rho}\nabla E_R = \kappa_P \rho c (a_RT^4-E_R)\\[3ex]
        \frac{\partial \mathbf{B}}{\partial t} = \nabla \times \left(\mathbf{v}\times\mathbf{B} \right) \\
    \end{cases}
\end{equation}

Here $\rho$ is the gas density, $\mathbf{v}$ the gas velocity, and $\mathbf{B}$ the magnetic field. Furthermore, $P_t$ is the total pressure, including both thermal and magnetic pressure, giving $P_t = P + \frac{B^2}{2}$, and $P$ is given by an ideal equation of state 
\begin{equation}
    P = (\gamma - 1) \rho u
\end{equation}
where $u$ is the internal energy, that is given by $u = \frac{k_B T}{(\gamma-1)\mu m_p}$, with $k_B$ being the Boltzmann constant, $\mu$ the mean molecular weight of the gas, $m_p$ the proton mass, and $T$ the temperature of the gas. The adiabatic index $\gamma$ is taken to be $1.42$ and the mean molecular weight $\mu$ is taken to be $\mu=2.353$ because the gas is considered a mix of hydrogen and helium in agreement with solar abundance \citep{Flock13, Szulagyi16, Flock17}.
$\Phi(\mathbf{r})$ is the gravitational potential due to the central star and the orbiting planet, i.e. $\Phi=-\frac{GM_{\star}}{r}  -\frac{GM_{p}}{\sqrt{|\mathbf{r}-\mathbf{r_p}|^2 + \epsilon^2}}$, where $\mathbf{r_p}$ is the position of the planet and $\epsilon$ is the gravitational softening length. $E_t = \rho u + \rho \frac{v^2}{2} + \frac{B^2}{2}$ is the total energy of the gas, including thermal, kinematic, and magnetic energy, but not including radiation energy, which is called $E_R$. 

The equations for radiative transfer are written following the formalism by \cite{Mihalas84} and make use of the Planck opacity $\kappa_P$ and the Rosseland opacity $\kappa_R$ \citep{Kley89, Commercon11}. $E_R$ is the radiation energy, $a_R = \frac{4\sigma_b}{c}$ the radiation constant, and $\lambda$ is a flux limiter chosen so that the radiation diffusion approaches the black body radiation in the optically thin scenario, while $\lambda \to 1/3$ in optically thick parts \citep{Levermore81}. In particular, in our setups

\begin{equation}
    \lambda = \frac{2+R}{6+3R+R^2}
\end{equation}
with $R=\frac{|\nabla E_R|}{\rho\kappa_RE_R}$. This way, $R$ gets very small in the optically thick scenario, giving $\lambda=1/3$, while $R$ gets very large when the opacity is low, meaning that $\lambda\sim\frac{1}{R}\sim \frac{\rho\kappa_RE_R}{|\nabla E_R|}$, and the diffusion term becomes 
\begin{equation}
    \nabla\frac{c\lambda}{k_R\rho}\nabla E_R \to \nabla (cE_R)
\end{equation} that approaches the black-body radiation, as in thermal equilibrium $E_R=\frac{B(T)}{c}=\frac{4\sigma_bT^4}{c}$. \citep{Kley89, Commercon11,Flock16,Szulagyi16}. 

In our case, non-ideal MHD effects have not been considered as, for example, no resistivity has been included. Nevertheless, these effects may be important when considering low-ionized gas like the one of protoplanetary discs, then we briefly discuss possible implementations and consequences of non-ideal mechanisms in Section \ref{sec:discussion}.

\subsection{Numerical methods}\label{sec:numerical_methods}

For all runs, the equations were solved via the PLUTO code \citep{Mignone07} with the HLLD Riemann solver \citep{Miyoshi05}, parabolic reconstruction \citep{Mignone14} and second order Runge-Kutta time integrator. In the MHD framework, we used the Constrained Transport module to ensure $\nabla \cdot \mathbf{B}=0$ \citep{Balsara99, Mignone07}. The radiative transfer equations were not solved in the same Godunov formalism, but an implicit solution was implemented in the code in order to be able to use reasonable timestep lengths.
In particular, the radiative transfer equations were solved by a module implemented in the PLUTO and described by \cite{Flock13}.

The PLUTO code was modified consistently by adding a module able to solve the implicit radiative transfer equations \citep{Flock13, Flock16, Flock17}. A new variable was implemented, i.e. the radiation energy $E_R$, that at the initial state follows $E_R= \frac{4\sigma_bT^4}{c}$. For simplicity, the Planck and the Rosseland mean opacities have been considered equal, with their values calculated as a function of pressure and temperature according to the opacity table by \cite{Zhu09}, with a floor value of $10\,cm^2g^{-1}$. The radiative transfer solver of PLUTO has been validated by comparing its results to the results of the code JUPITER \citep{Szulagyi16}, which includes a radiative transfer solver as well. More details and plots are provided in Appendix \ref{sec:PLUTO_JUPITER}.
In order to keep the runs stable, a floor density equal to $7\times 10^{-9} M_{\odot}/AU^3$ ($10^{-6}$ in code units) was implemented.

The computational grid domain always stretches logarithmically from 2.08 AU to 13 AU in the radial direction, then linearly from $0$ to $2\pi$ in azimuth and linearly from $\sim0.436\pi$ to $\pi/2$ in co-latitude, so that only the upper half of the disc is simulated and the lower part is just considered to be symmetric and mirrored in the analysis. This gives a grid opening angle of $\sim 0.2$ radians. This range turned out to be large enough in capturing the meridional circulation flow. In fact, in all the runs presented in Section \ref{sec:results}, the gas density at upper latitudes is comparable to the floor density, i.e. 3 or 4 orders of magnitude lower than the gas density in the mid-plane.

In the low-resolution runs, we had 210x22x720 ($r$, $\theta$, $\phi$) cells (44 effective cells in the co-latitude direction when mirrored), while high-resolution runs had 420x44x1440 cells. The Courant number has been chosen to be $0.25$. Units have been chosen to have $G=1$, then $M_{\odot}$ has been used for mass, $5.2 AU$, i.e. Jupiter's semi-major axis, for length, and consequently, $\frac{P_J}{2\pi}$ for time, where $P_J$ is the orbital period of Jupiter.

\subsection{Boundary conditions}\label{sec:boundary_conditions}

In the azimuthal direction, the boundary condition is simply periodic. The boundary condition in the co-latitude direction is symmetric at mid-plane, meaning that scalar quantities ($\rho$, $P$, $E$), parallel velocities, and perpendicular magnetic fields simply reflect with the same sign in the ghost cells, while perpendicular velocities and parallel magnetic fields reflect but switching sign. At the inner co-latitude boundary, that in these coordinates corresponds to the higher latitudes, the values in the ghost cells are equal to the value in the first active cell, with the caveat that the velocity in the $\theta$-direction is set to $0$ if positive, i.e. if gas is being injected in the grid domain. The radiation energy is chosen so that the radiation temperature, which is defined as $T_R^4 = \frac{cE_R}{4\sigma_b}$, is set to a value of $10K$ in the ghost cells to allow cooling of the upper disc layers. The radial and azimuthal components of magnetic fields in ghost cells are set equal to the first active cell, while the $\theta$-component is calculated to keep $\nabla\cdot\mathbf{B}=0$.

Radially, similar prescriptions are applied. All ghost-cell quantities have the same value as the domain cells except the radial velocity, which is set to 0 whenever the gas radial velocity points towards the domain, in order to prevent inflow, and the radial magnetic field, which is automatically calculated to keep it divergence-free. This time though, the radiation temperature is set to $10K$ only if the gas temperature is lower than that value, otherwise, in ghost cells, we set $T_R=T$.

\subsection{Disc model}\label{sec:disc_model}

First of all, the central star is taken to be a Sun-like star, with $M_\star=M_{\odot}$, $R_\star=R_{\odot}$, and $T_\star=T_{\odot}$ on the surface. The surface density profile of the disc has the form $\Sigma = \Sigma_0 (R/R_0)^{-1/2}$, with $\Sigma_0$ calculated so that the total mass of the disc at the beginning is $M=0.01M_{\odot}$. The vertical density profile follows
\begin{equation}
    \rho(z,R) = \rho_0(R)\exp\left(-\frac{z^2}{2H^2} \right)
\end{equation}
with $H=hR$ and the aspect ratio $h=0.05$ being constant. Here $R=r\sin(\theta)$ and $z=r\cos(\theta)$ are the correspondent cylindrical coordinates, with $R_0$ being any reference radius, for example, the semi-major axis of the forming planet.

We initialize our setups with a vertical constant temperature, while the radial profile is calculated to ensure a constant aspect ratio $h$ throughout the disc. This brings to a profile $T\propto R^{-1}$. The initial velocities are set so that the radial and the co-latitude components are set to $0$, while the azimuthal component is set to be equal to the Keplerian velocity $v_{\phi} = \sqrt{\frac{GM_{\odot}}{R}}$.

The initial magnetic field is assigned to be poloidal, i.e. $B_r=B_\phi=0$, while $B_\theta = -B_0 \left(\frac{R_0}{R}\right)$, where $B_0$ is calculated so that a given value for the magnetic-$\beta$, i.e. $\beta=\frac{P}{P_{\mathrm{mag}}}=\frac{2P}{B^2}$, is obtained at planet location. The components of the magnetic field $\mathbf{B}$ are assigned via a magnetic potential $\mathbf{A}$, defined so that $\mathbf{B} = \nabla \times \mathbf{A}$, that ensures that $\nabla \cdot \mathbf{B} = 0$. In particular, in this case, $A_r = A_\theta = 0$, and $A_\phi = \frac{B_0}{\sin(\theta)}$. This configuration ensures also that the disc does not have magnetic forces at the beginning of the simulation, since $\nabla \times \mathbf{B}=0$.

This setup was run for 50 orbits with no magnetic fields so that a new thermal equilibrium is reached under the heating and cooling effects. Then, every simulation with or without magnetic fields was run starting from this relaxed initial condition, which we identify as time 0, meaning that the first relaxation with no magnetic fields happens between $t=-50$ orbits and $t=0$. In all setups, the simulation ran then for 200 Jupiter-orbits without the planet reaching a steady state with MHD (or HD) + radiative effects. Then, a Jupiter-like planet is gradually grown at $5.2$ AU over 150 orbits, and it was assumed to be in a circular orbit. The planet was a point-mass with the gravitational potential according to Equation \ref{eq:global}, with a softening length equal to the cell size at the planet location, meaning $0.04AU$ in low-resolution and $0.02AU$ in high resolution. The mass of the planet is grown via a continuous function as
\begin{equation}
    M_p(t) = M_J \cdot \begin{cases}
    0\;\;\;\;&-50<t<200\\
    \sin \left(\frac{\pi}{2}\frac{t-200}{150}\right)^2\;\;\;\;&200\le t < 350\\
    1\;\;\;\;&t\ge350\\
    \end{cases}
\end{equation}
where the time $t$ is in units of Jupiter-orbits here. Since the planet is implemented as a point mass kept artificially in a circular orbit, it does not migrate nor change its eccentricity or inclination. The simulation was then run for another 250 orbits to let the disc relax and reach a new equilibrium with the planet.

\section{Results}\label{sec:results}

Both low- and high-resolution simulations were carried out. A low resolution was used because it allowed exploring quickly many different configurations, while only a few selected setups were run in a high-resolution scheme to resolve potential MRI effects that are resolution-dependent. On average, a low-resolution MHD simulation took about 300'000 core-hours (about 2'500 node-hours on our machines), while high-resolution runs took as much as 4'000'000 core-hours (30'000 node-hours). In our case, the latter ones translated to about 20-day runs, meaning that even improving the resolution by a factor of 2 in each dimension, we would have had setups that need several months to be completed. This is why we chose (420x44x1440) to be our highest resolution. 
In order to start our simulations from the same relaxed setup, a disc with no magnetic fields and no planet was run for 50 Jupiter orbits, in order to allow the structure to relax and settle. All the runs that are going to be mentioned in this section were started from the result of this relaxation run, with the proper addition of the initial magnetic fields in the MHD cases.

\subsection{Initial Conditions and Equilibrium}\label{sec:IC}

\begin{figure}
\includegraphics[width=\columnwidth]{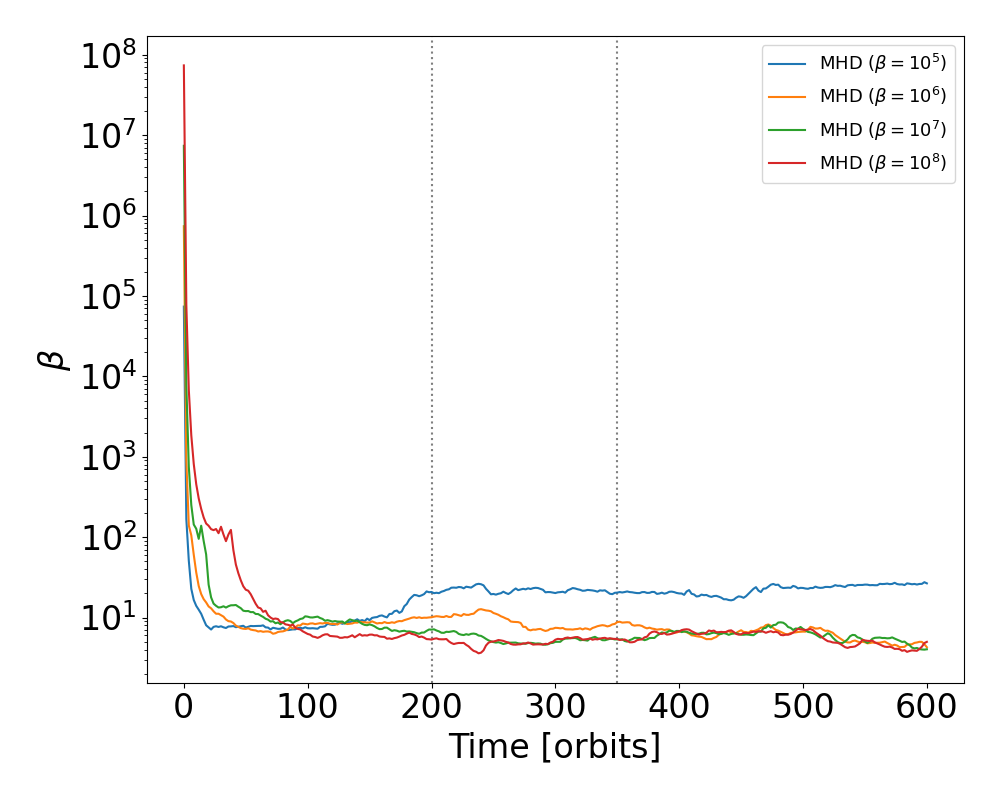}
\caption{Globally integrated $\beta$ (see Equation \ref{eq:beta}) in four different simulations with various initial magnetic field strengths ($\beta$ parameter): $10^5,10^6,10^7,10^8$. Despite the initial value, all simulations' $\beta$ parameters evolve to the same value, but on different timescales. The grey dotted lines represent the moments when the planet is injected into the simulation and when the planet completes its growth and the mass-point becomes 1 Jupiter-mass.}\label{fig:magnetic_energy}
\end{figure}

Before analyzing the meridional circulation and gap formation, we first examined the effect of the initial conditions on the final disc structure.
In order to do that, we run a set of 4 different low-resolution MHD simulations, each of them starting with a different initial value of the parameter $\beta$, i.e. $\beta = 10^5, 10^6, 10^7, 10^8$ at planet location. This has been done by choosing the initial value of $B_0$, described in Section \ref{sec:methods}, accordingly. These runs showed that the initial magnetic field magnitude is not relevant in terms of the final outcome of the system. In fact, even though the total magnetic energy of the disc had different initial values in the four cases, it rapidly grew to the same magnetic field strength value over the first 100 orbits, as shown in Figure \ref{fig:magnetic_energy}. After that, this value remained roughly constant at a global $\beta \sim 10$ for the entire simulation, even after the planet is injected and fully grown.  The only difference we see is the timescale of the magnetic energy growth, which is longer in the cases with a larger initial $\beta$. The fast growth of magnetic fields and magnetic energy is due to the upwind term in the induction equation for $\frac{\partial \mathbf{B}}{\partial t}$ (see Equation \ref{eq:global}). In fact, the fast orbital velocities quickly generate a strong toroidal field starting from the vertical-field configuration. In this and the following examples, a global $\beta$ parameter of the disc was defined and calculated as
\begin{equation}\label{eq:beta}
    \beta = \frac{2\int P dV}{\int B^2 dV}
\end{equation}
where the integrals were done over the whole disc volume $V$. This global parameter is the ratio of the total thermal and magnetic energy in the disc, similarly to the local $\beta$, which is the ratio of the local thermal and magnetic pressure. It also happens to be approximately equal to the local $\beta$ at the planet's location at the beginning of the simulation, while the two differ at the end of the runs when most of the magnetic energy is found at higher latitudes (see plots in Section \ref{sec:gap_opening}).

\begin{figure}
\includegraphics[width=\columnwidth]{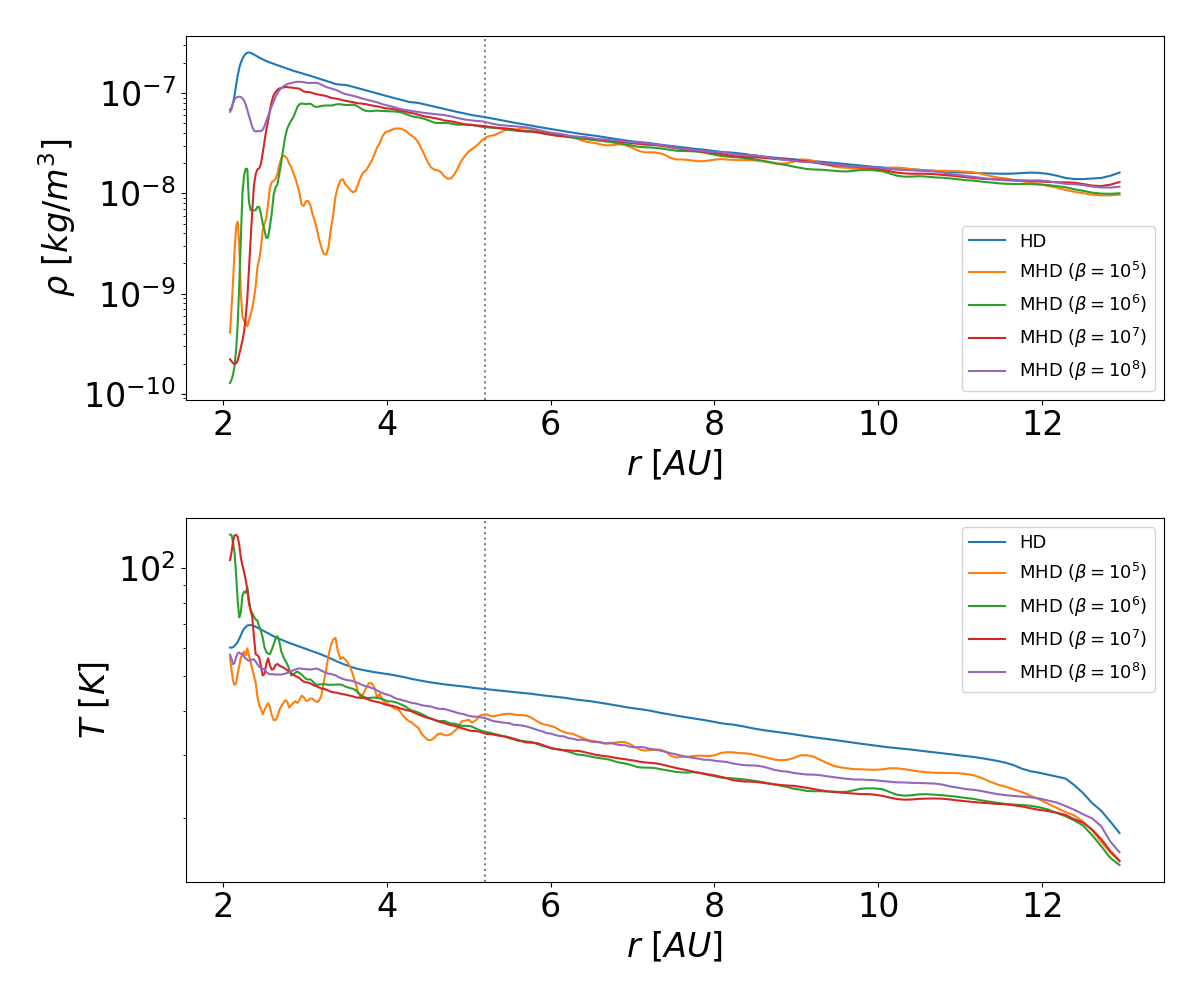}
\caption{Density and temperature profiles taken at the planet's position ($\theta=\pi/2$, $\phi=0$) of five different simulations (one HD and four MHD with $\beta = 10^5,10^6,10^7,10^8$) after 200 orbits, before the planet is introduced. Higher $\beta$'s correspond to a smoother evolution and smaller perturbations, especially around the planet formation location at $5.2 AU$. The grey dotted line represents the planet's location.}\label{fig:main_comparison_100_mag}
\end{figure}

We found that the longer growing timescale for the magnetic field strength keeps the disc structure stable and prevents large perturbations in the density and temperature profiles, which could affect the results. In fact, the case with the initial $\beta=10^5$ proved to be particularly subject to perturbations across the whole disc, that modify the dynamics at the planet location. On the other hand, higher initial $\beta$'s showed to produce much smoother results and to preserve the structure of the disc around the planet much better (Figure \ref{fig:main_comparison_100_mag}). The lowest initial $\beta$ that produced unquestionably stable and smooth results is $\beta=10^7$ and that is why this value has been chosen as the initial condition for most of the MHD runs. Nevertheless, the case with initial $\beta=10^5$ will be mostly analyzed in Appendix \ref{sec:beta_1e5}.

\begin{figure}
\includegraphics[width=\columnwidth]{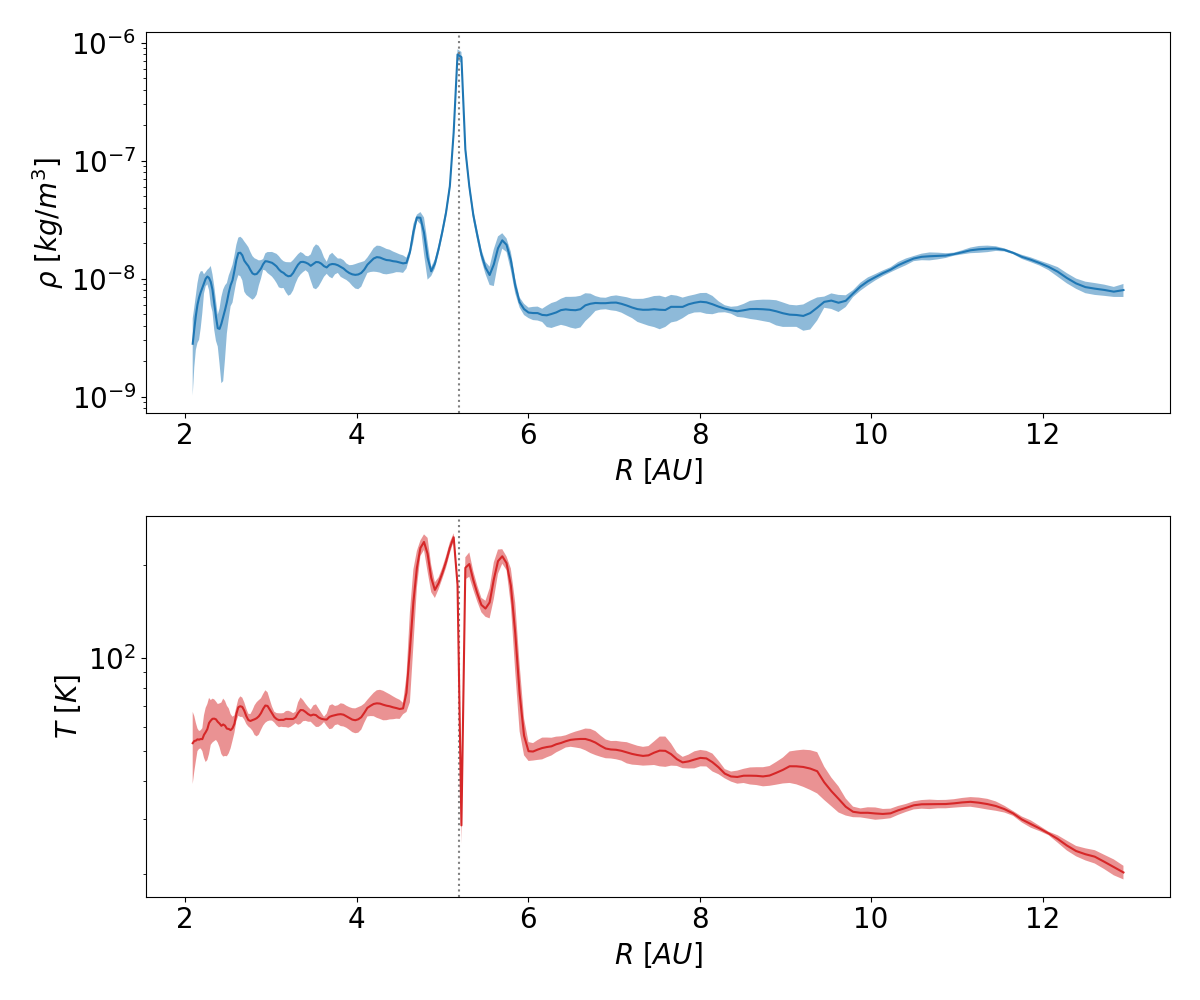}
\caption{Density and temperature profiles taken at the planet's position ($\theta=\pi/2$, $\phi=0$) in the MHD case with initial $\beta=10^7$. The solid lines represent the average profiles from 560 to 600 orbits, while the colored areas represent the standard deviation of the profiles. The grey dotted line represents the planet's location.}\label{fig:MHD_steady_state}
\end{figure}

Every simulation was run for 200 orbits without the planet, then the planet itself took 150 orbits to grow via a \textit{mass-taper} function, and another 350 orbits are carried out to let the disc stabilize and reach the steady state. Even though in cases with low viscosity many thousands of orbits are needed in order to reach perfect equilibrium \citep{Hammer18}, in our case, a few hundred orbits turn out to be enough to reach a sufficiently steady state. In order to check that, we verified that the density and temperature profile did not change significantly over time at the end of the simulation. In particular, Figure \ref{fig:MHD_steady_state} shows the average density and temperature profiles of the disc between 560 and 600 orbits and their standard deviation. It is clear that, except for some smaller fluctuations through the disc due to MHD effects, both the density and temperature profiles around the planet are very stable. This suggests that the disc structure after 600 orbits can be taken as a final configuration, and we do not expect it to change significantly even though the same runs were run for longer.

\subsection{Meridional Circulation}\label{sec:meridional_circulation}

\begin{figure*}
\includegraphics[width=0.95\textwidth]{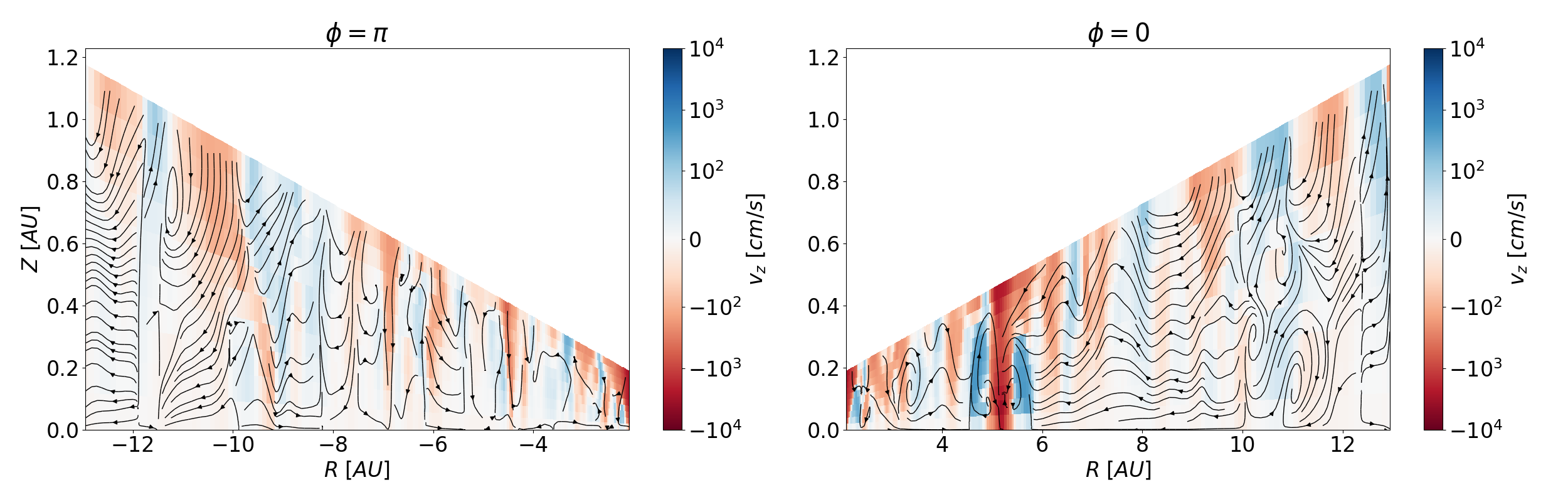}
\caption{Two vertical sections of the protoplanetary disc after 300 orbits, with no magnetic fields, when the planet has already fully formed and the disc has already fully relaxed. The sections are taken at $\phi=0$ and $\phi=\pi$ with respect to the position of the planet, meaning that one section corresponds to the planet's location, and the other is the opposite. The arrows represent the direction of gas velocity, while the color scale represents the value of the vertical velocity component, whether it is positive or negative. Only the upper hemisphere is shown.}\label{fig:HD_meridional_circulation}
\end{figure*}

\begin{figure*}
\includegraphics[width=0.95\textwidth]{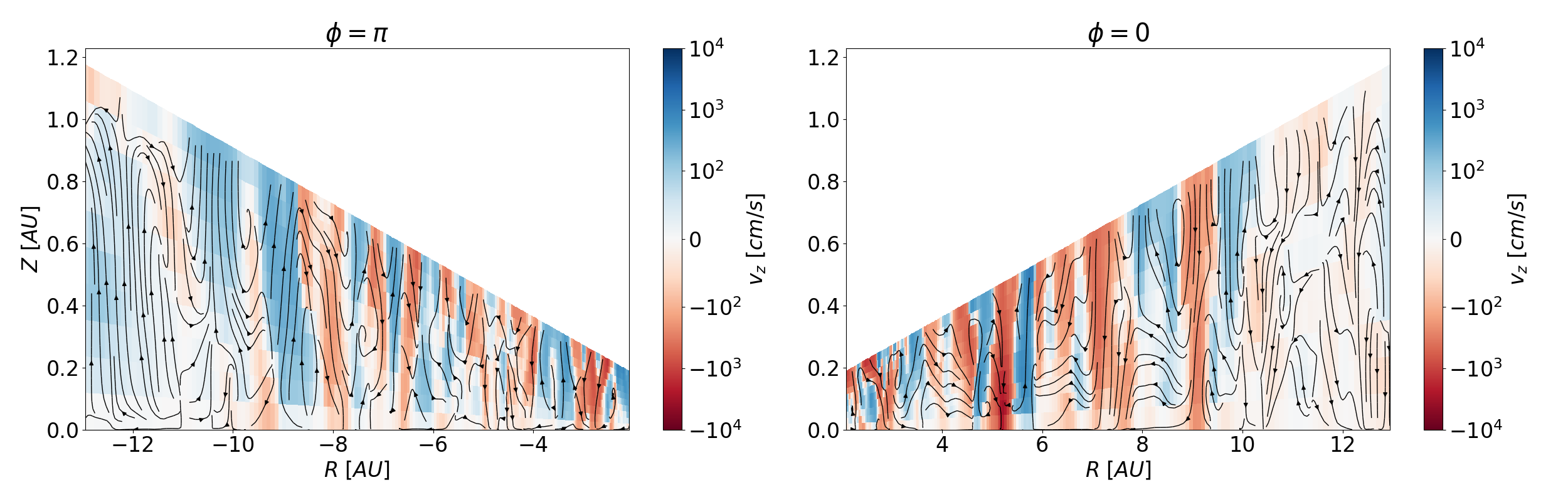}
\caption{Two vertical sections of the protoplanetary disc after 300 orbits in an MHD simulation ($\beta=10^7$), when the planet has already fully formed and the disc has already fully relaxed. The sections are taken at $\phi=0$ and $\phi=\pi$ with respect to the position of the planet, meaning that one section corresponds to the planet's location, and the other is the opposite. The arrows represent the direction of gas velocity, while the color scale represents the value of the vertical velocity component, whether it is positive or negative. Only the upper hemisphere is shown.}\label{fig:MHD_meridional_circulation}
\end{figure*}

\begin{figure*}
\includegraphics[width=0.42\textwidth]{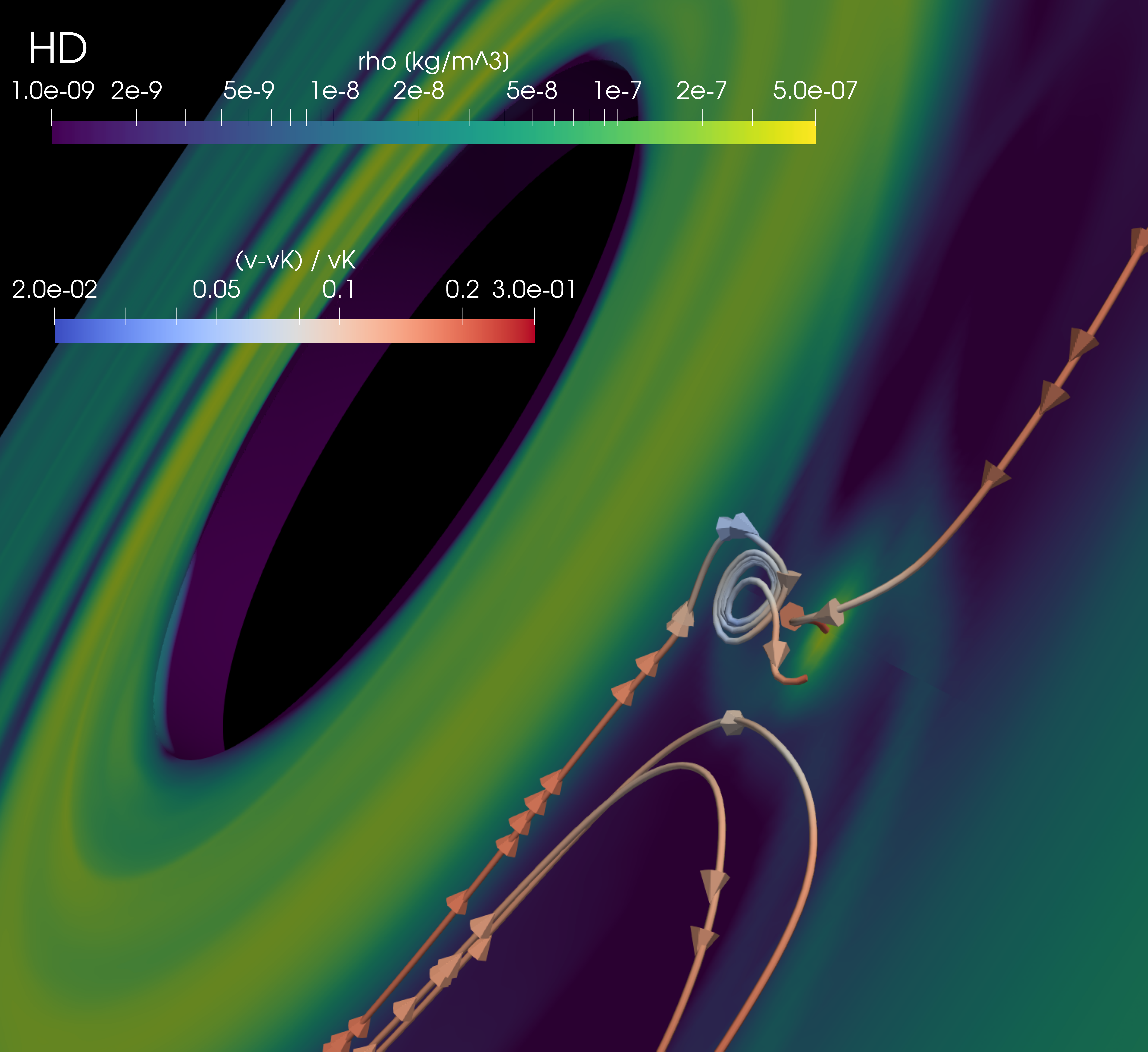}
\includegraphics[width=0.42\textwidth]{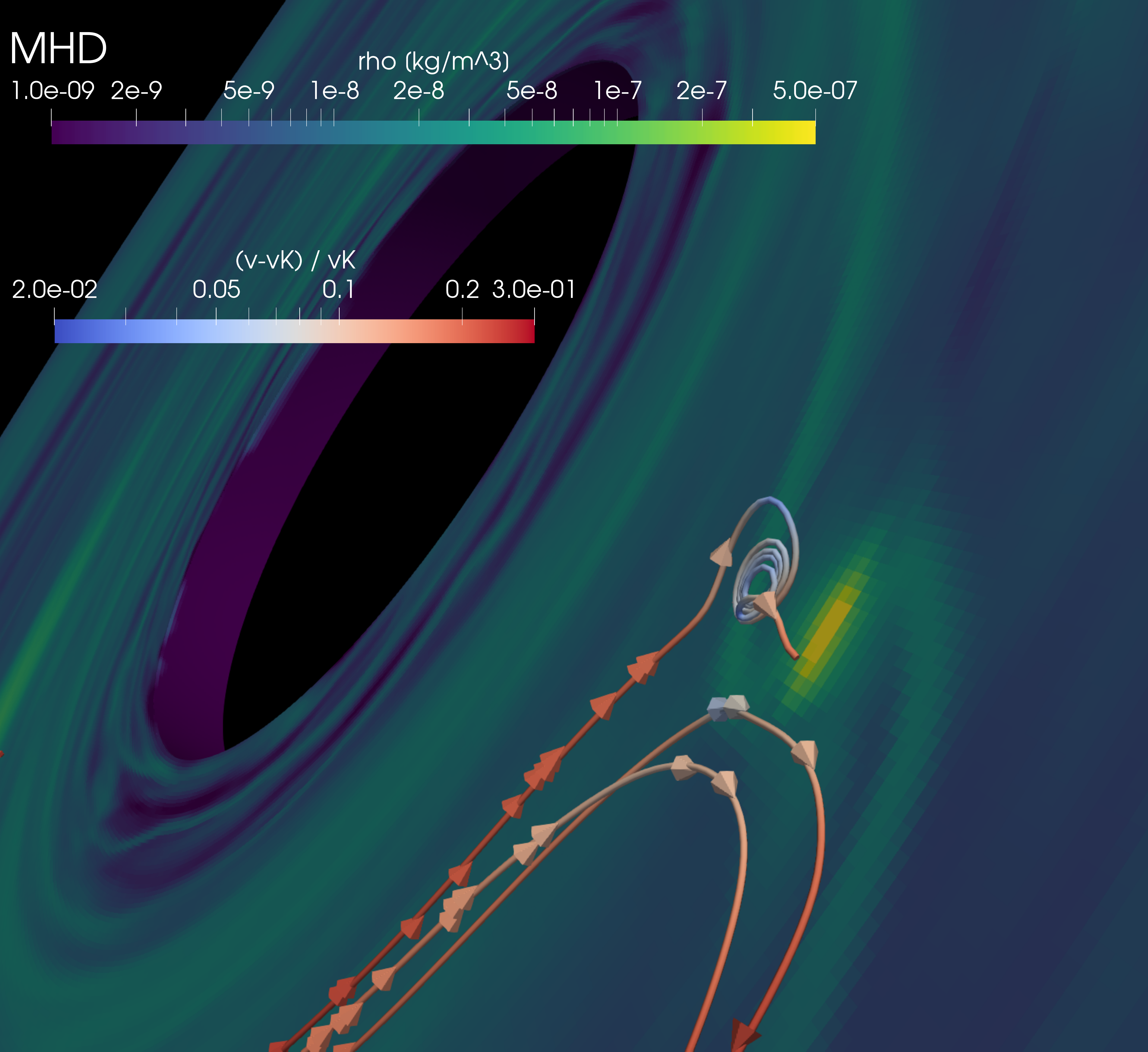}
\caption{3D streamlines of the accretion flow towards the planet in the radiative HD (left) and MHD with initial $\beta=10^7$ (right) cases. The background colormap represents the volume density in the mid-plane. The color of the streamlines shows the magnitude of the difference between the gas velocity and the planetary Keplerian velocity in units of the planetary Keplerian velocity itself}.\label{fig:paraview}
\end{figure*}

When a Jupiter-like planet forms in a protoplanetary disc, it is expected to open up a gap, while its spiral wakes cause a meridional circulation of material to launch. While the spiral wakes themselves bring material up and down in the entire circumstellar disc, the downward flow is the strongest above the planet causing the vertical accretion stream \citep{Szulagyi14,Szulagyi21}. Figure \ref{fig:HD_meridional_circulation} shows two sections of the protoplanetary disc, one (right side) correspondent to the planet's location, the other (left side) on the  opposite side in the protoplanetary disc after 600 orbits. The color scale corresponds to the vertical velocity component, showing the strong upward and downward flows in the entire circumstellar disc driven by the spiral wakes of the planet, but with the highest magnitude just above the planet. In addition, the gas streamlines are overplotted as well, showing the vertical circular motions, i.e. the meridional circulation. 

Figure \ref{fig:MHD_meridional_circulation} shows the same as Fig. \ref{fig:HD_meridional_circulation} , but in the MHD case with an initial $\beta=10^7$. In this case, too, the meridional circulation pattern and the velocity magnitude around the planet appear to be very similar to the HD case, but in other areas of the circumstellar disc further away from the planet, the vertical velocities and circulation are much stronger than in the HD simulation. In this case, naturally, the entire circumstellar disc is more turbulent.

We followed the accretion flow down to the planet, with the help of 3D velocity streamlines representing the true gas motion (Figure \ref{fig:paraview}). The left panel is the HD simulation, while the right one is the MHD simulation with initial $\beta=10^7$. The color map represents the volume density at the mid-plane, where the planet is located on the right of the plot, while the streamline's colors represent the velocity magnitude in the rotating frame normalized to the Keplerian velocity. The 3D streamlines are above the mid-plane, and show how the flow is going toward the planet from the vertical direction. As the streamlines in the planetary gap encounter the spiral wake of the planet (which is a shock-front), the gas slows down losing angular momentum, Therefore those streamlines spiral down to the circumplanetary disc, some directly hitting the planet's polar regions and thus being accreted by the planet. Most of the gas entering the Hill-sphere will not be accreted by the planet, the streamlines in the mid-plane region of the CPD spiral outward and bring away gas from the planet region back to the circumstellar disc. Such recycling effect is observed regularly in disc-planet simulations, recently again described by \cite{mol21}.

All in all, the accretion flow does not change whether magnetic effects are included or not, highlighting that the flow dynamics are really driven by the planetary spiral arms. Between the two panels in Figure \ref{fig:paraview}, the density values are the main difference and that is the reason why the accretion rate and Hill sphere's masses are so different, as described in Section \ref{sec:accretion_CPD}.

\begin{figure*}
\includegraphics[width=0.49\textwidth]{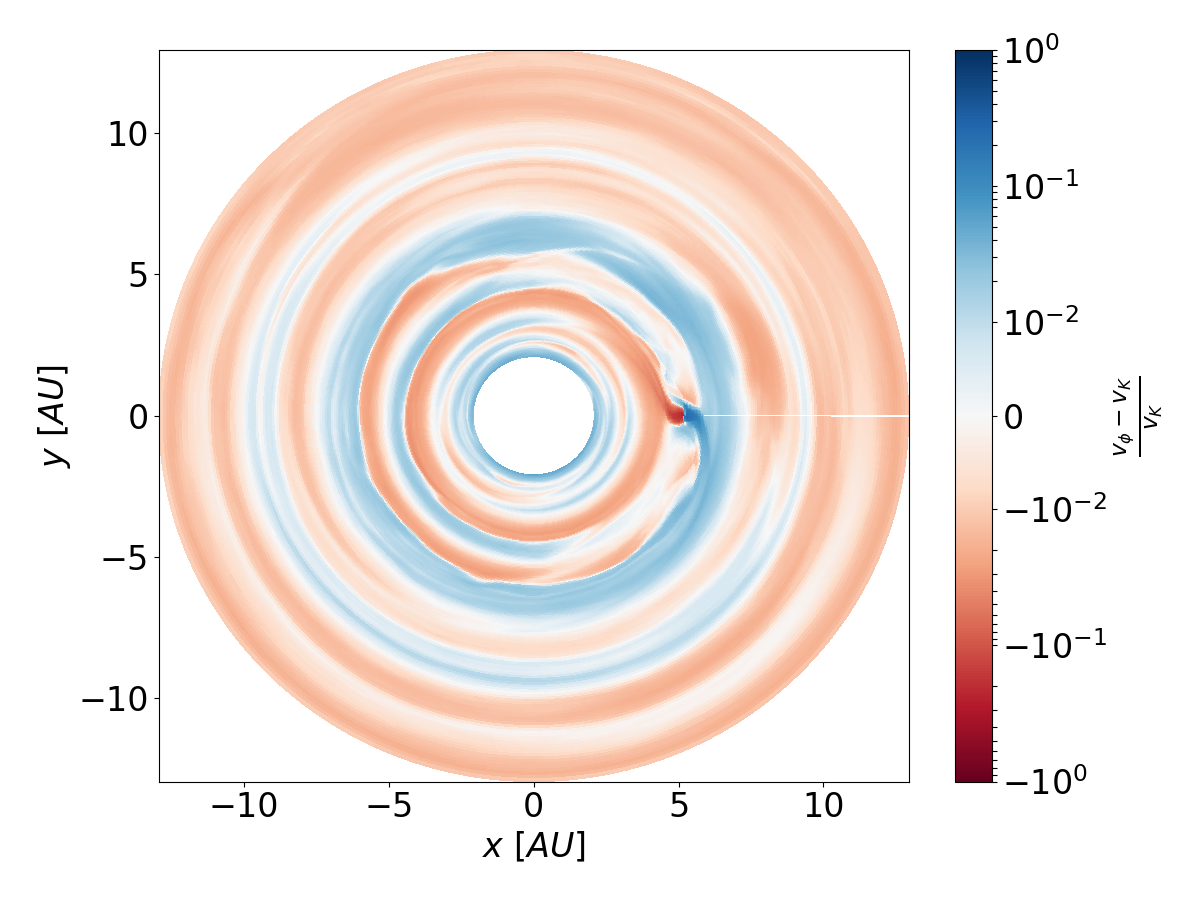}
\includegraphics[width=0.49\textwidth]{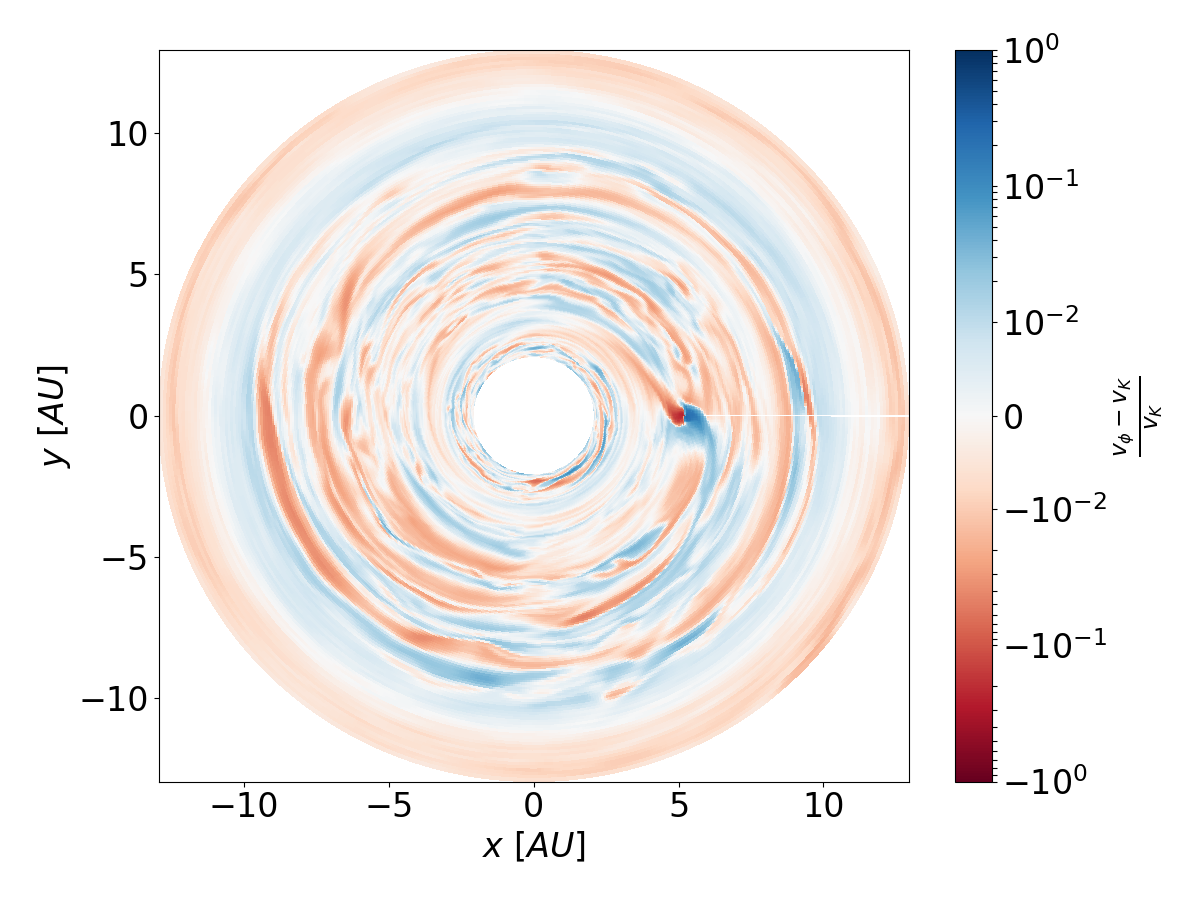}
\caption{The plots show the difference between the local Keplerian velocity and the actual azimuthal velocity, normalized to the Keplerian velocity itself. Red areas represent sub-Keplerian motion, while blue ones represent the super-Keplerian flow. The left panel shows the HD case, while the right panel shows the MHD case (initial $\beta=10^7$). The Keplerian regions (white) are dust-trapping regions, but because the MHD case is more turbulent, dust-trapping in rings is more difficult. }\label{fig:vel_diff}
\end{figure*}

We also examined the azimuthal velocity in the mid-plane, normalized by the local Keplerian velocity, to compare the HD and MHD cases, see Figure \ref{fig:vel_diff}. In the HD case, Figure \ref{fig:vel_diff} (left panel) shows clearly the super and sub-Keplerian regions due to the pressure rise and fall at the outer and inner gap edges. These mostly axisymmetric structures are strongly disturbed in the MHD case (see the right panel in Figure \ref{fig:vel_diff}), which looks much more turbulent and less axisymmetric. Which part of the mid-plane is sub-Keplerian or super-Keplerian is important in the context of dust trapping, because the Keplerian areas are particularly subject to dust accumulation \citep{Kato10}. From the plots, it is clear that in the absence of magnetic fields, the disc structure is more regular, with rings of sub- and super-Keplerian material easily identified. On the other hand, in the MHD case, the disc structure is again much more turbulent, as expected. Rings are not easy to define since they do not have clear boundaries nor shapes. Furthermore, such a fragmented velocity profile would continuously change with time, while gas is orbiting around the star with different Keplerian velocities. We can infer then that dust accumulation should be much more difficult in the ideal MHD case compared to the HD one and, consequently, formation mechanisms such as the streaming instability \citep{Youdin05} are much more difficult to happen.

\subsection{Effective viscosity}

\begin{figure}
\includegraphics[width=\columnwidth]{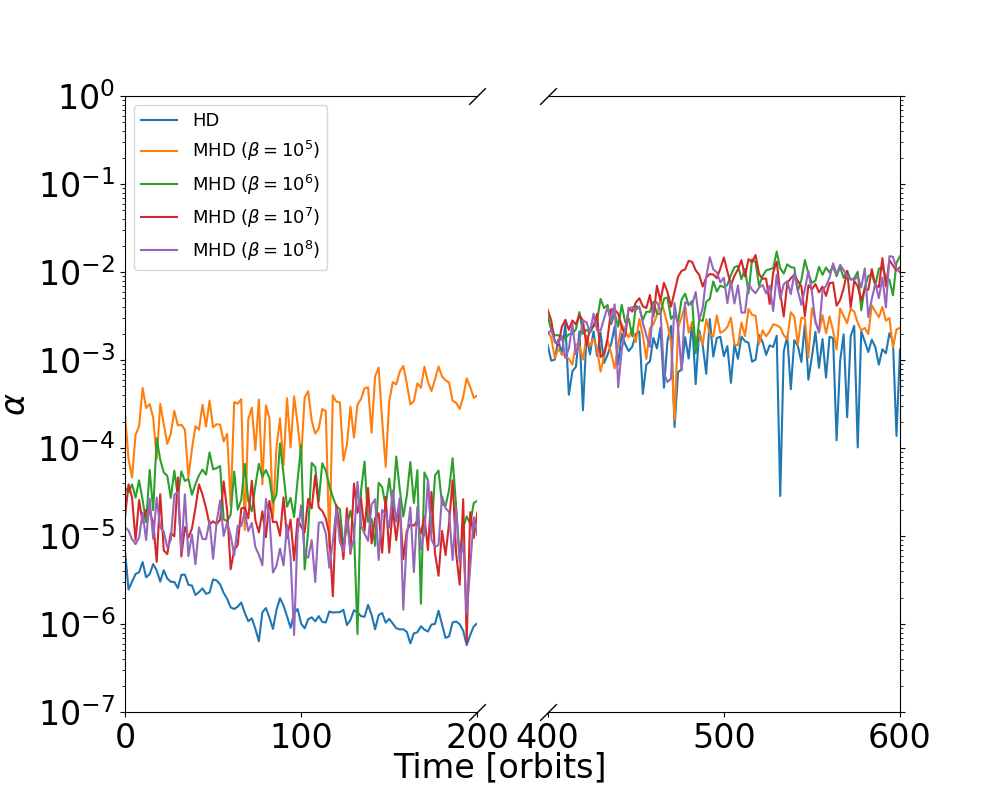}
\caption{Effective $\alpha_R$ parameter calculated via the Reynolds stress, in the HD case compared to the four MHD cases, with initial $\beta=10^5,10^6,10^7,10^8$, tracked up to the formation of the planet at 200 orbits, and then once the planet is completely formed from 400 to 600 orbits.}\label{fig:alphav_comparison}
\end{figure}

The mixing due to the meridional circulation and (mainly) the encounter with the spiral shock wake of the planet have an effect also on the viscous behavior of the gas. On one hand, the circular motions generate vortexes on the lower scales, which then cause the growth of turbulence, that can be described by a turbulent, or effective, viscosity in the disc, that can be quantified via an $\alpha$ parameter \citep{Shakura73}. Furthermore, as the gas parcels encounter the spiral shock fronts of the spiral wake of the planet, they lose angular momentum and drive accretion as well. This can be translated into an effective viscosity as well, given the angular momentum loss mechanism.

We estimated the effective viscosity of the circumstellar disc by calculating the Reynolds stress as shown by \cite{Flock11,Flock13}. The $r-\phi$ component of the Reynolds tensor is defined as $T_R = \rho v'_\phi v'_r$, where $v' = v - <v>$, and $<>$ is hereafter the average over time. The corresponding $\alpha_R$ parameter, i.e. relative to velocity turbulence or to the Reynolds stress, is then calculated as
\begin{equation}
    \alpha_R = \frac{\int \frac{T_R}{c_s^2}dV }{\int \rho dV}
\end{equation}
where the integral is over the whole disc domain.
Figure \ref{fig:alphav_comparison} shows the evolution of $\alpha_R$ in the HD and MHD cases, with different initial $\beta$'s, for the first 200 orbits, before the planet starts forming, and then once the planet is completely formed from 400 to 600 orbits. The plot shows that the kinematic effective viscosity is lowest for the HD case, i.e. $\alpha_R \simeq 10^{-6}$, while the magnetized cases have a higher viscosity, between $10^{-3}$ and $10^{-5}$, as specified in Table \ref{tab:alpha_R}. This viscosity acts also as a heating mechanism, balancing radiative cooling.

\begin{figure}
\includegraphics[width=\columnwidth]{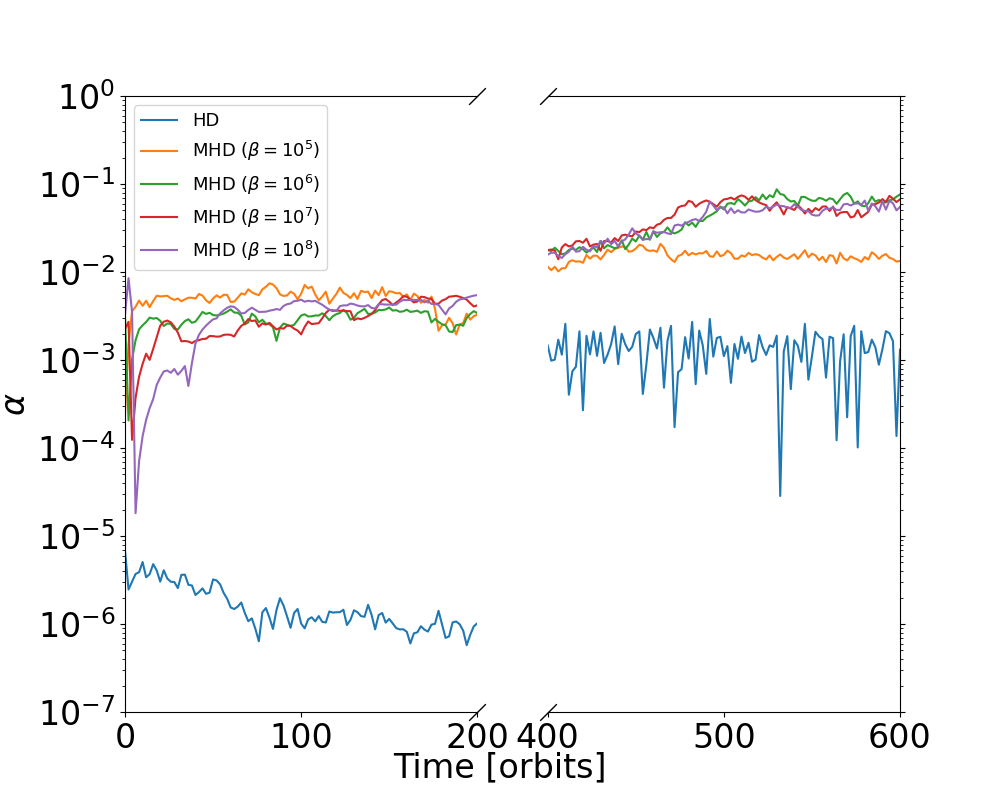}
\caption{Effective $\alpha$-viscosity parameter in the HD case compared to the four MHD cases, with initial $\beta=10^5,10^6,10^7,10^8$, tracked up to the formation of the planet at 200 orbits, and then once the planet is completely formed, from 400 to 600 orbits.}\label{fig:alpha_comparison}
\end{figure}

A similar estimate was already done by \cite{Stoll17}, which found that the presence of a planet up to $100M_\oplus$ did not affect much the turbulence in the disc in the case of 3D HD inviscid and vertically isothermal simulations.
In our case, when a Jupiter-like planet is present in the disc, it clearly dominates turbulent viscosity through the spiral wakes. This causes the effective $\alpha$ to grow up to values between $10^{-3}$ in the HD case and $10^{-2.2}$ in the MHD cases. This means that in the HD case, the spiral wakes contribute to three orders of magnitude higher alpha in the entire disc, meaning that the effects of spirals are the main angular momentum loss mechanism in the disc. In our research field, since the MRI is no longer considered the main angular momentum loss mechanism, there has been an ongoing search for other processes that can remove angular momentum, for example, magnetic winds \citep{Gressel15, Bethune17} and various instabilities like the vertical shear instability \citep{Stoll14, Barker15}, the zombie wave instability \citep{Marcu15, Marcus16}, Rossby-wave instability \citep{Lovelace99} and non-ideal MHD effects \citep{Bai11, Bai13}. The spiral shock wakes of planets were not really considered before, or mainly just in the CPD \citep{Szulagyi14, Zhu16, SzulagyiMordasini17}, or without quantification of the effective alpha \citep{Szulagyi21}. In the MHD simulations (Table \ref{tab:alpha_R}), the  $\alpha_R$ values jump to higher final values than in the HD, but in these MHD runs the initial (first 200 orbits without the planet) $\alpha_R$ was also much higher than in the HD case (Fig. \ref{fig:alpha_comparison}). One can conclude that even in the MHD case also the planet spirals contribute to the effective viscosity.

\begin{table}
	\centering
	\caption{Calculated values of $\alpha_R$ before (from the beginning to 200 orbits) and after (from 400 to 600 orbits) the planet is fully formed.}
	\label{tab:alpha_R}
	\begin{tabular}{lll} 
		\hline
		setup & $\log_{10}(\alpha_R)$ before & $\log_{10}(\alpha_R)$ after \\
		\hline
		HD                 & $-5.9 \pm 0.1$ & $-2.9 \pm 0.2$ \\
		MHD ($\beta=10^5$) & $-3.5 \pm 0.2$ & $-2.7 \pm 0.1$ \\
		MHD ($\beta=10^6$) & $-4.5 \pm 0.2$ & $-2.2 \pm 0.1$ \\
		MHD ($\beta=10^7$) & $-4.9 \pm 0.2$ & $-2.2 \pm 0.1$ \\
		MHD ($\beta=10^8$) & $-5.0 \pm 0.0$ & $-2.4 \pm 0.1$ \\
		\hline
	\end{tabular}
\end{table}

Velocity fluctuations and material circulation are not the only sources of turbulence in the disc though. In fact, following again \cite{Flock11,Flock13}, magnetic fields can give their contribution to the effective viscosity via the Maxwell stress, calculated from the $r-\phi$ component of the Maxwell tensor, i.e. $T_M = -\frac{B_\phi B_r}{4 \pi}$, where the full value of the magnetic field is taken into account. In this case, the magnetic equivalent viscosity parameter is calculated as 
\begin{equation}
    \alpha_M = \frac{\int \frac{T_M}{c_s^2}dV }{\int \rho dV}
\end{equation}
and the total effective viscosity is estimated as $\alpha = \alpha_R + \alpha_M$. The evolution of the total viscosity is shown in Figure \ref{fig:alpha_comparison}. Now, $\alpha$ is much larger in magnetized discs, reaching values between $10^{-3}$ and $10^{-2}$ before the planet is injected and up to $10^{-1}$ after that, compared to values around $10^{-6}$ and $10^{-3}$ respectively in HD simulations (see Table \ref{tab:alpha_tot}). Once again, the presence of the planet deeply influences the global viscosity of the disc because of the turbulence and magnetic fields developing in the planetary spiral wakes. We should expect a much more viscous behaviour in MHD cases then, that can influence momentum transport and gap opening, even though we do not expect viscous heating to be much larger in the magnetic case in the bulk of the circumstellar disc, except the very inner regions close to the star.

\begin{table}
	\centering
	\caption{Calculated values of $\alpha=\alpha_R + \alpha_M$ before (from the beginning to 200 orbits) and after (from 400 to 600 orbits) the planet is fully formed.}
	\label{tab:alpha_tot}
	\begin{tabular}{lll} 
		\hline
		setup & $\log_{10}(\alpha)$ before & $\log_{10}(\alpha)$ after \\
		\hline
		HD                 & $-5.9 \pm 0.1$ & $-2.9 \pm 0.2$ \\
		MHD ($\beta=10^5$) & $-2.3 \pm 0.1$ & $-1.8 \pm 0.1$ \\
		MHD ($\beta=10^6$) & $-2.5 \pm 0.1$ & $-1.3 \pm 0.1$ \\
		MHD ($\beta=10^7$) & $-2.6 \pm 0.1$ & $-1.3 \pm 0.1$ \\
		MHD ($\beta=10^8$) & $-2.4 \pm 0.2$ & $-1.3 \pm 0.1$ \\
		\hline
	\end{tabular}
\end{table}

In fact, when the Maxwell stress tensor was defined, the total value of magnetic fields was considered, instead of only its fluctuations, as in the case of the Reynolds stress. As described by \cite{Flock11, Flock13, Flock17}, the mean magnetic field strength acts as a momentum transport mechanism, equivalently to viscosity. At the same time, fluctuations in the magnetic field structure underline the presence of instabilities, such as the Magneto-Rotational Instability (MRI, \citealt{Balbus98}), that translates into turbulence and into viscous heating as well. In general, we can then distinguish between a mean Maxwell tensor $T_{M,\mathrm{mean}} = -\frac{<B_\phi>< B_r>}{4 \pi}$ and a residual or turbulent tensor $T_{M,\mathrm{res}} = T_{M} - T_{M,\mathrm{mean}}$ where again $T_{M} = -\frac{B_\phi B_r}{4 \pi}$. These tensor components can be used to identify relative viscosity parameters $\alpha_{M,\mathrm{mean}}$ and $\alpha_{M,\mathrm{res}}$, that sum up to retrieve the total equivalent viscosity $\alpha = \alpha_R + \alpha_{M,\mathrm{mean}} + \alpha_{M,\mathrm{res}}$. Trivially, $\alpha_R$ is the only contribution in HD simulations, while the three components can have different importance in MHD simulations. In particular, in our MHD models, the dominant component is by far the one due to mean magnetic fields. In fact, the Reynolds component is found to be a couple of orders of magnitude lower, and the residual magnetic fields give a negligible equivalent $\alpha$ from the turbulent Maxwell stress tensor. This means that, at least in this resolution, magnetized discs are expected to be much more viscous than HD setups, but, at the same time, viscous heating should not substantially kick in.

\subsection{Gap opening}\label{sec:gap_opening}

\begin{figure*}
\includegraphics[width=0.7\textwidth]{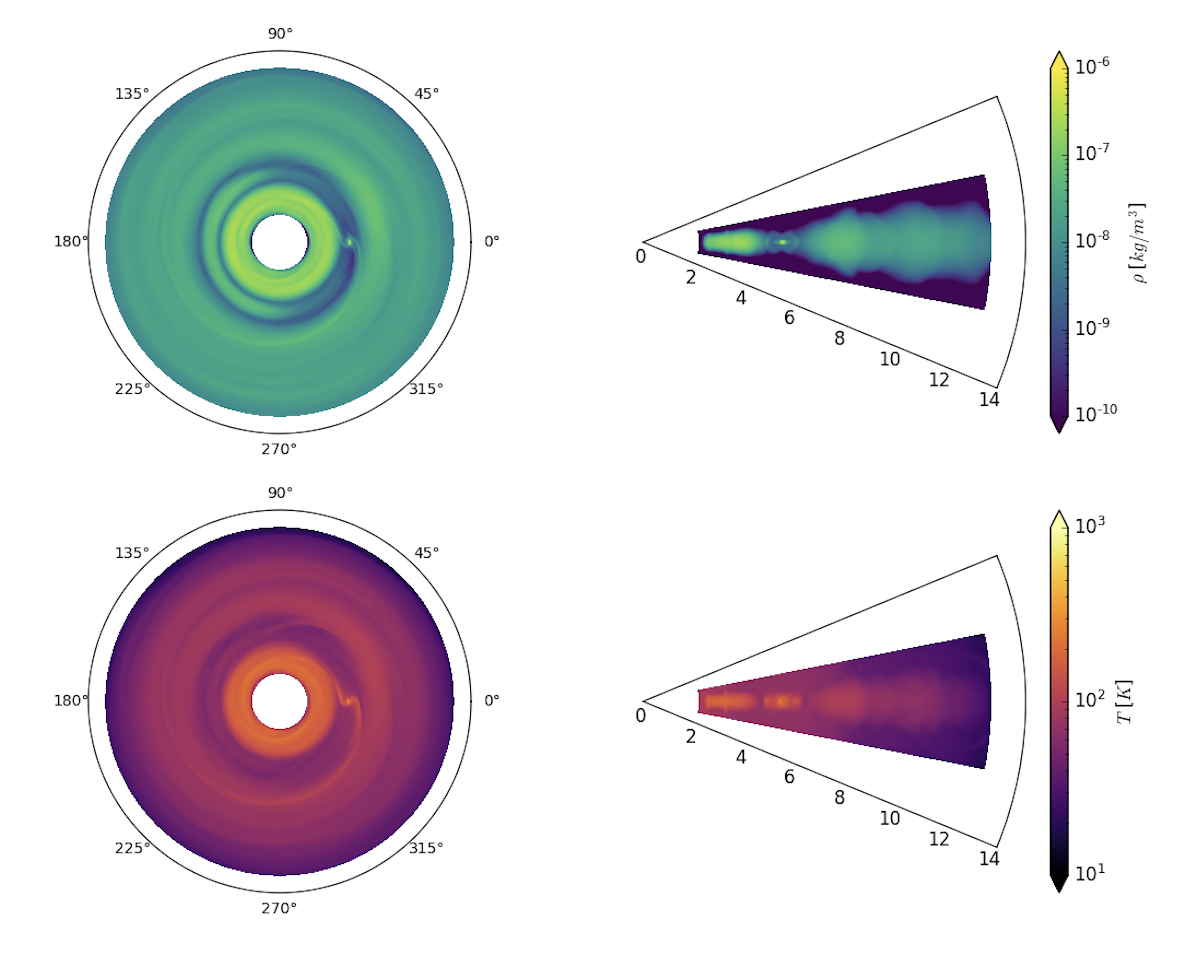}
\caption{Section of the disc at the end of the HD run. On the left panel, the  mid-plane sections are shown, while on the right panel the vertical sections at the planet location. The top row shows the density and the bottom row shows the temperature.}\label{fig:main_2D_HD}
\end{figure*}

After having analyzed the meridional circulation of gas, in this section we address the issue of gap opening, comparing the HD and the MHD cases. First of all, without magnetic fields, a gap is quite clearly opened in the disc. This is observable in Figure \ref{fig:main_2D_HD} where density and temperature profiles of the HD case are shown. In the 2D sections, the gap is visible around the planetary orbit and presents a lower density, other than a lower temperature. Because of the gap formation, gas piles up in the inner and outer disc, increasing the density, especially at the inner boundary.  At the planet location, we can also see a concentration of material, in both figures. The resolution is not high enough to determine the structure of such a clump, but that could be in general a trace of the formation of a circumplanetary disc, given the peak in both density and temperature. Since we expect the disc to slowly reach a definitive equilibrium configuration after thousand of orbits \citep{Hammer18}, it is likely that all the remaining gas in the gap will eventually clear out.

\begin{figure*}
\includegraphics[width=0.7\textwidth]{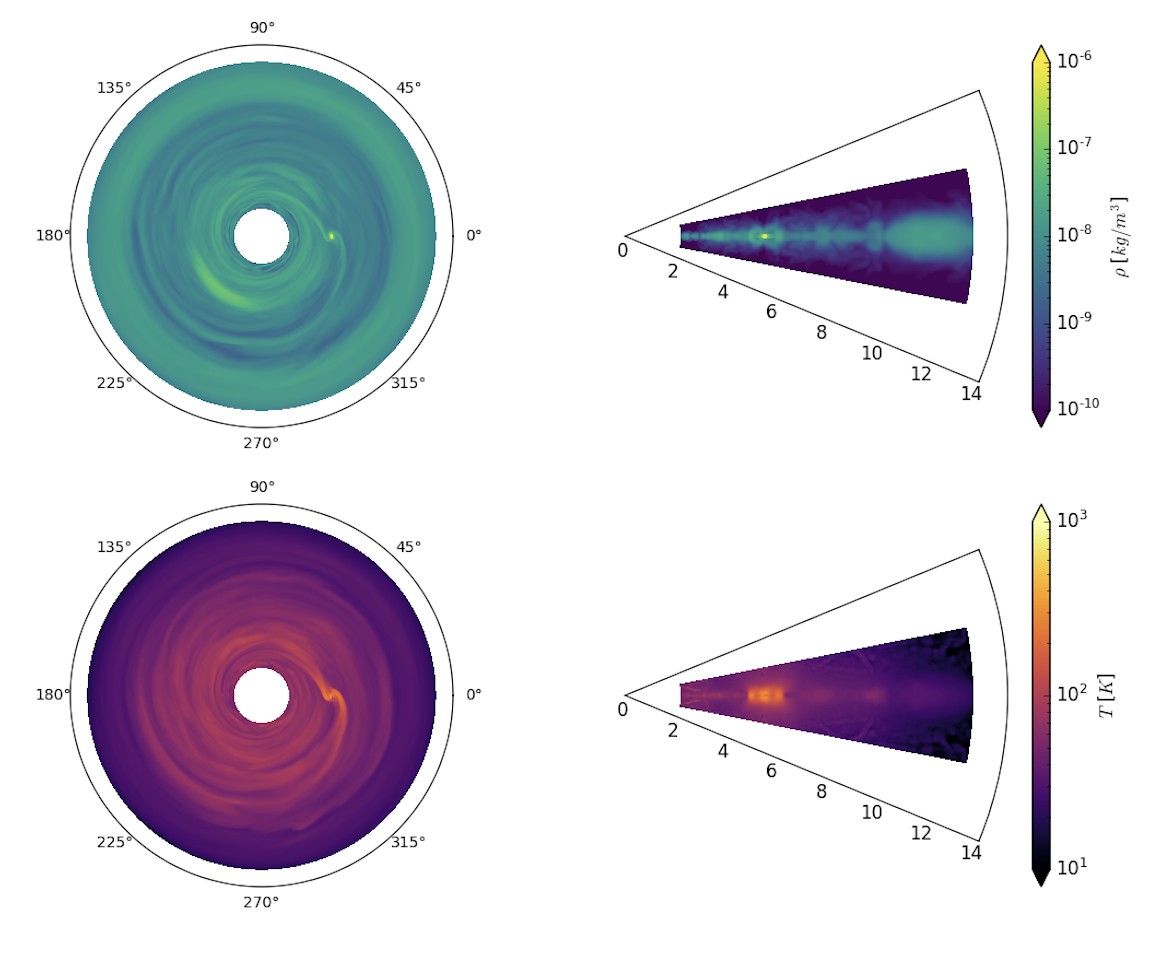}
\caption{Section of the disc at the end of the MHD run with initial $\beta=10^7$. On the left mid-plane sections are shown, while on the right the vertical sections are at the planet's location. The top row shows density and the bottom row shows temperature. }\label{fig:main_2D_MHD}
\end{figure*}

The same section and quantities have been measured in the MHD case and are shown in Figure \ref{fig:main_2D_MHD}, with a snapshot of the mid-plane and vertical sections taken after 600 orbits, with a fully-formed planet. In this case, the presence of a gap is not clearly spotted. The density and temperature distributions reveal a much more turbulent structure, as already mentioned in Section \ref{sec:meridional_circulation}, and there is no strong decrease in temperature or density around the planetary orbit, even though circumplanetary material is again visible right around the planet. This again shows that the gap formation and the CPD formation are not necessarily linked, but there can be circumplanetary material (either an envelope or a disc) even when a planetary gap is not present or is yet to form.

\begin{figure}
\includegraphics[width=\columnwidth]{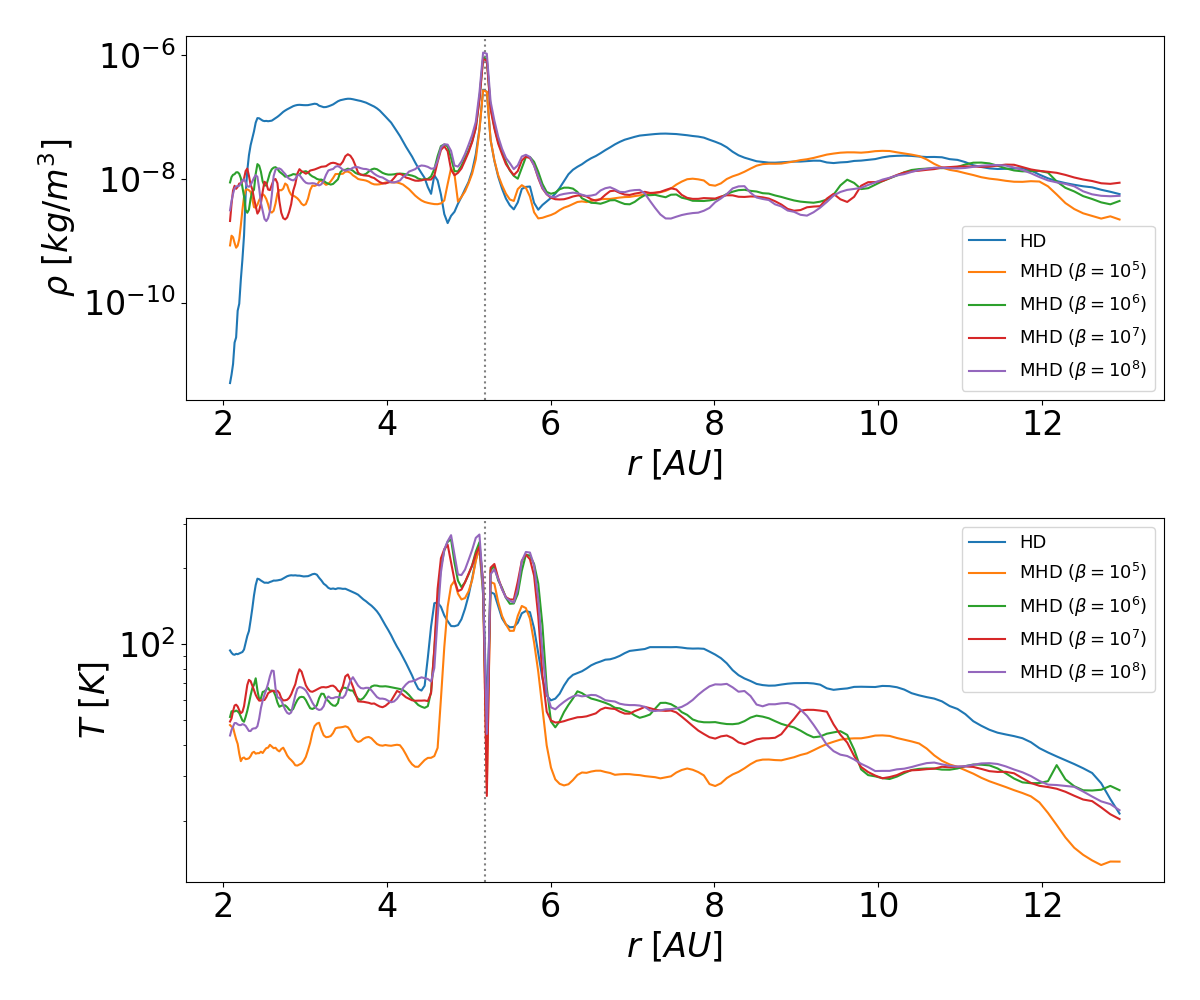}
\caption{Comparison of the mid-plane temperature and density profile at the planet longitude ($\phi=0$). The HD result is compared to four different MHD runs with different initial values for the magnetic $\beta$ ($10^5, 10^6, 10^7, 10^8$). The grey dotted line represents the planet's location.}\label{fig:main_comparison_300}
\end{figure}

More details are shown in Figure \ref{fig:main_comparison_300}. These plots compare the mid-plane density (top) and temperature (bottom) profiles at the planet's location of the HD simulation with the four MHD setups (initial $\beta=10^5, 10^6, 10^7, 10^8$). In the HD case, the gap structure is clearly visible in both density and temperature, with a decrease of more than one order of magnitude in the first case. At the same time, the peaks in density and temperature exactly at the planet's location reveal the presence of circumplanetary material, and the formation of the gap forces most of the gas to pile up in the outer and especially in the inner part of the circumstellar disc, enhancing density close to the inner radial boundary.

On the other hand, the MHD runs show a completely different behavior. First of all, the case with $\beta=10^5$ reveals itself again as a more perturbed case, due to the faster growth of magnetic fields and their larger effects on the circumstellar disc structure. Nevertheless, the other runs show, as expected, very similar profiles. The gap is mostly absent, as there is not any clear decrease in the density distribution. The circumplanetary material is again present, with a similar density profile, but higher in magnitude, suggesting more massive CPDs. Even the pile-up of materials in the inner protoplanetary disc is not detected, as a consequence of the lack of a deep gap. At the same time, the temperature profile resembles closely the HD results in the vicinity of the planet, with a small positive offset, while the disc appears to be much colder in the rest of its structure.

\begin{figure}
\includegraphics[width=\columnwidth]{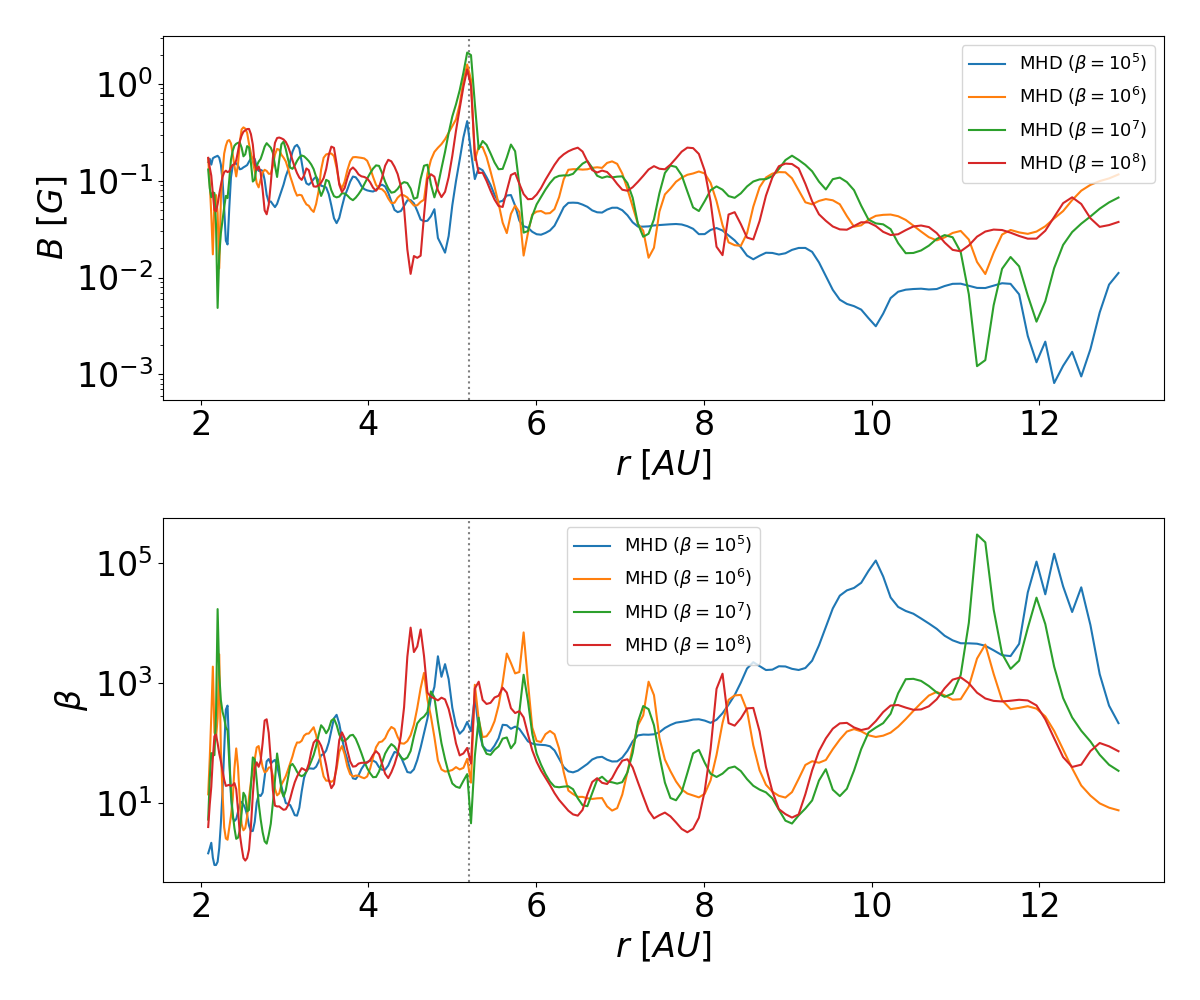}
\caption{Comparison of the mid-plane magnetic field magnitude and $\beta=\frac{2P}{B^2}$ at the planet longitude ($\phi=0$). Four different MHD runs with different initial values for the magnetic $\beta$ ($10^5, 10^6, 10^7, 10^8$) are compared, showing a magnetic field peak inside the gap, and a $\beta$ down to the value of $10$. The grey dotted line represents the planet's  location.}\label{fig:magnetic_comparison_300}
\end{figure}

The lack of a gap opening in MHD is somehow expected from the results of Section \ref{sec:meridional_circulation}. In fact, the conditions for a gap to open can be analytically estimated by comparing the timescales of the different mechanisms acting in the gap formation process, i.e. the gravitational torque and the viscous timescales \citep{Lin80, Papaloizou84}. In this context, \cite{Armitage10} showed that, following \cite{Takeuchi96}, a gap is expected to open when the following condition is met
\begin{equation}
    50 \alpha h^2 m^{-1} < 1
\end{equation}
where $\alpha$ is the viscosity parameter, $h$ the aspect ratio, and $m = M_p / M_\star$ the mass ratio between the planet and the star. In our case, with $h=0.05$ and $m=10^{-3}$, the conditions yields to $\alpha < 8 \times 10^{-3}$. During a simulation $h$ can decrease down to $\sim 0.035$ (see Figure \ref{fig:aspect_ratio}), giving the condition $\alpha < 1.5 \times 10^{-2}$. Since the equivalent viscosity before the planet is injected is about $10^{-6}$ in HD runs, this condition is fully met and a gap is allowed to open. On the other hand, an equivalent $\alpha$ between $10^{-3}$ and $10^{-2}$, such as the one shown by MHD runs, would be exactly on the edge and would not allow the formation of a full gap, and it even grows at the end of the simulations to much higher values.
Moreover, the presence of strong magnetic fields in the gap would perturb the dynamics, providing an extra pressure term that would act against the opening of a gap. This effect is not dominant, as the $\beta$ parameter never gets below $10$, as shown in Figure \ref{fig:magnetic_comparison_300}, but at the same time not completely negligible.

\begin{figure}
\includegraphics[width=\columnwidth]{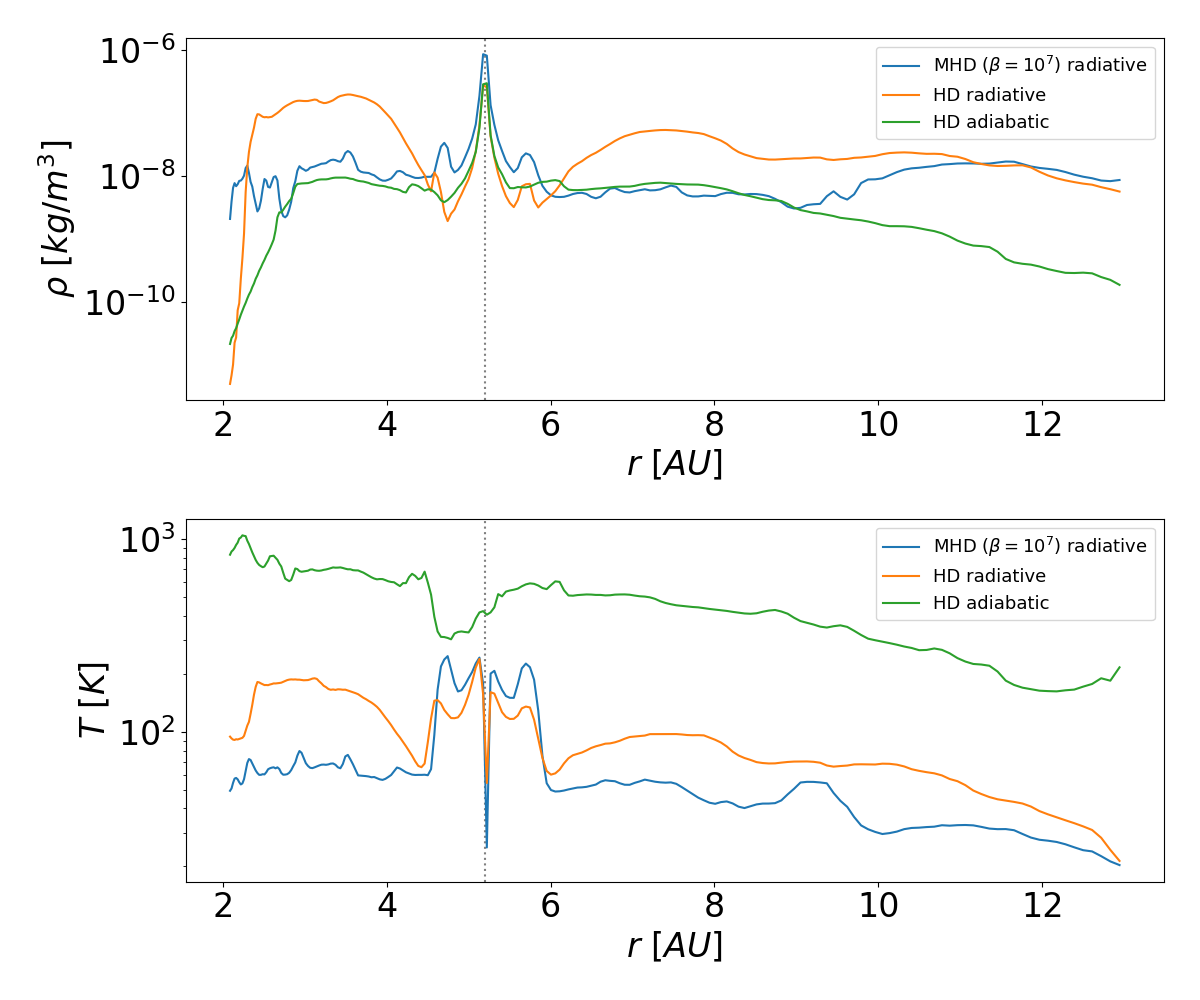}
\caption{Comparison of the mid-plane temperature and density profile at the planet longitude ($\phi=0$). The HD and MHD (initial $\beta=10^7$) results are compared to the same profiles of an adiabatic HD simulation. The grey dotted line represents the planet's location.}\label{fig:main_comparison_300_adiab}
\end{figure}

In fact, an extra pressure term can be quite important in controlling the mechanism of gap formation. The simpler way to study its effect is to enhance the thermal pressure in an HD disc. The faster way to do it is to deactivate radiative transfer and let the disc evolve adiabatically over time. This would prevent the disc to radiate energy away and would produce higher temperatures and, consequently, higher pressures, even though a proper equilibrium may not be reached if the disc keeps getting hotter over time. The result is shown in Figure \ref{fig:main_comparison_300_adiab}. The adiabatic run is actually much hotter than the others, meaning also that the thermal pressure is much higher in the disc. Consequently, the torque of the planet is not dominant in the gap-opening process anymore. This translates into a lack of gap, with a density profile very similar to the MHD case, even though slightly lower.

\begin{figure}
\includegraphics[width=\columnwidth]{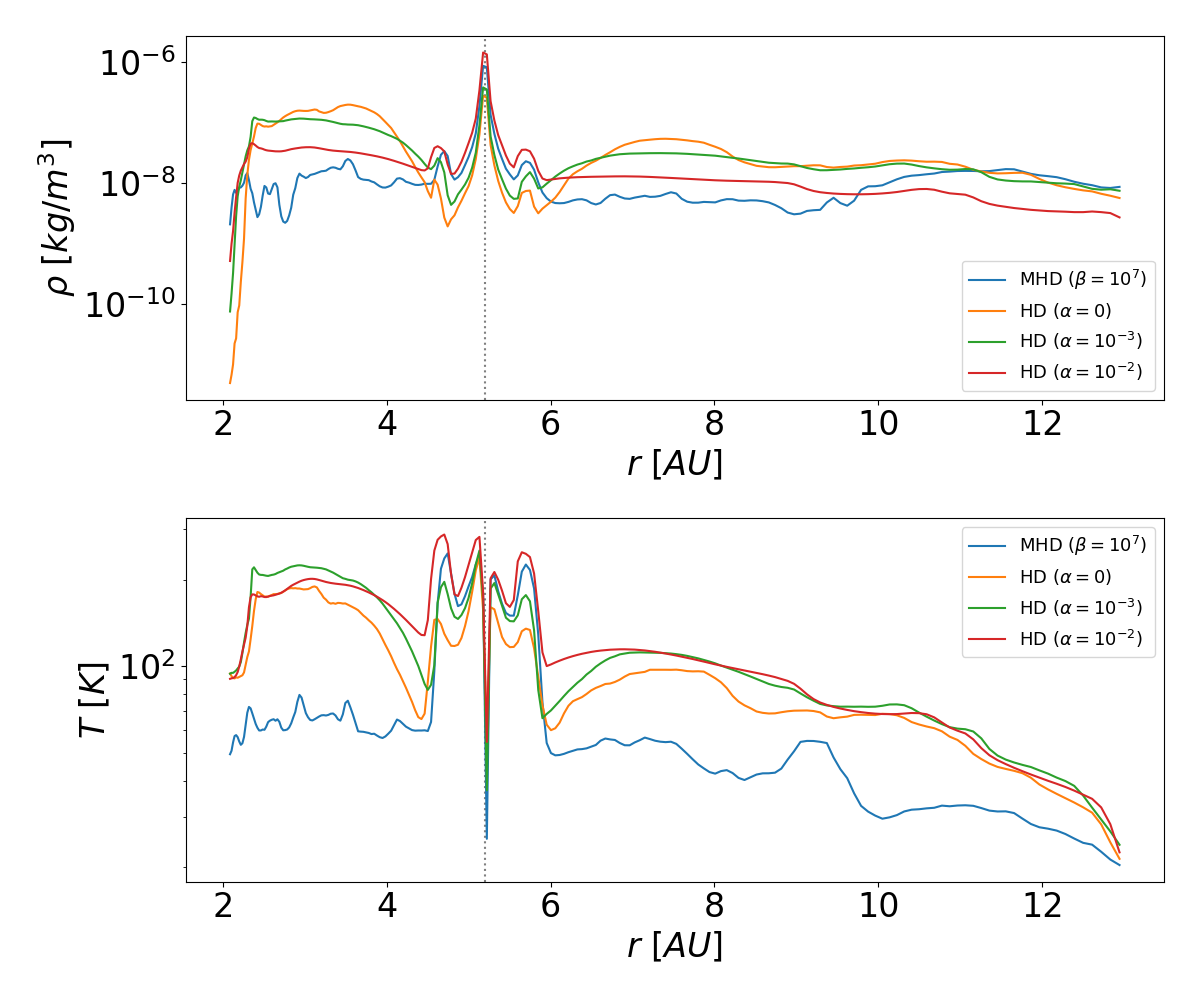}
\caption{The inviscous HD and the MHD case with initial $\beta=10^7$ are compared to two HD viscous setups ($\alpha=10^2,10^3$) in density (top) and temperature (bottom) after 600 orbits, when the planet is fully formed and the disc has relaxed. The grey dotted line represents the planet's location.}\label{fig:main_comparison_300_viscous}
\end{figure}

To elaborate more on the effective viscosity, even though the gap opening process is prevented by the high effective viscosity, an MHD simulation with an equivalent $\alpha$ is not directly comparable to a viscous HD setup. In fact, in the MHD case, the effective viscosity is dominated by mean magnetic fields, and consequently, we do not expect the temperature to be affected as much as it is in a viscous run. In order to prove that, we ran two HD simulations implementing viscosity, with artificial $\alpha$ parameters set to be $10^{-3}$ and $10^{-2}$, and the results are shown in Figure \ref{fig:main_comparison_300_viscous}. In the Figure, we observe that the density profile becomes similar to the MHD case (initial $\beta=10^7$) while $\alpha$ increases, even though the HD setups show a slightly higher density in the entire circumstellar disc. The thermodynamics is extremely different though. Despite the viscous behavior, the magnetized disc remains much colder than the HD viscous cases, especially far away from the planet. This confirms that the role of MHD can be compared to the effects of viscosity regarding the density distribution, while they are not quite similar with regard to the temperature of the disc.

\subsection{Thermodynamics comparison}\label{sec:thermodynamics_comparison}

All the previous results have been obtained in runs with radiative transfer, except the adiabatic comparison shown in Figure \ref{fig:main_comparison_300_adiab}. We also tested whether similar results would be obtained in case other strategies are implemented together with MHD. For example, using an isothermal equation of state would keep the disc temperature low and possibly damp any effective viscosity, changing in fact the meridional circulation pattern and the gap-opening process.

\begin{figure}
\includegraphics[width=\columnwidth]{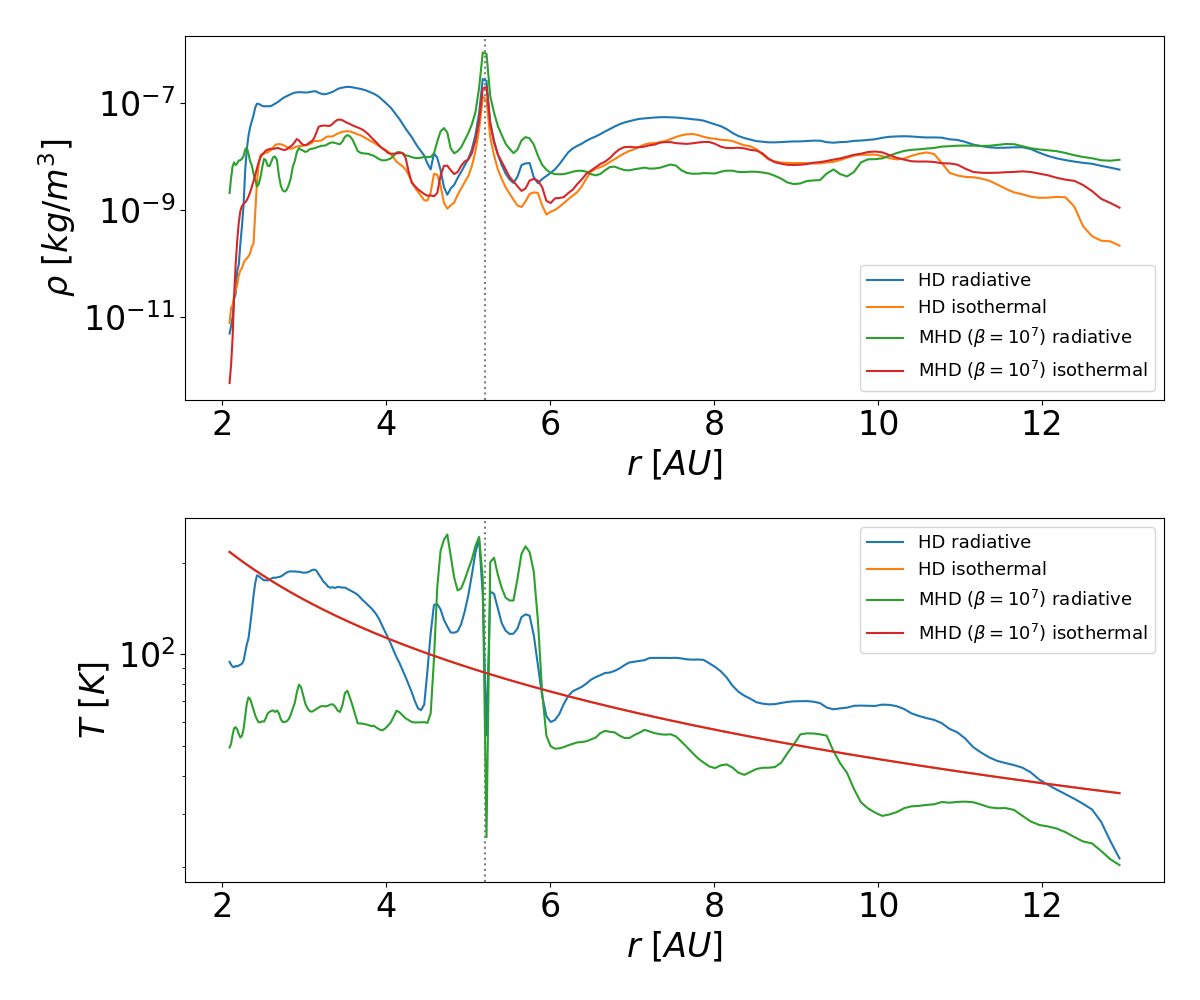}
\caption{The same HD and MHD (initial $\beta=10^7$) setups are compared in the two different cases (radiative and adiabatic) in both density and temperature. The grey dotted line represents the planet's location. In the isothermal cases, the temperature profiles overlap.}\label{fig:main_comparison_300_iso}
\end{figure}

As shown in Figure \ref{fig:main_comparison_300_iso}, in the HD case, an isothermal run produces only smaller differences compared to the radiative tests. In fact, we have a similar gap profile, even though the isothermal case is slightly more massive. On the other hand, the situation changes when comparing magnetized discs. In fact, the isothermal setup appears to have a much clearer gap, almost comparable to the HD cases. This means that the combination of an isothermal treatment with MHD does not produce the same effects of the MHD + radiative transfer implementation. The main difference is to be found in how the equivalent $\alpha$ evolves in the disc. In fact, in this case, the disc is not dominated by the mean Maxwell contribution anymore. This is found to be actually comparable to the contribution of the turbulent velocities (Reynolds stress) for most of the disc evolution. The total viscosity parameter is itself always around $10^{-5}$ before the planet is injected. This means that viscosity is much lower and produces the same behavior as in the HD case, i.e. a gap opening in the disc, a similar density profile, and a meridional circulation pattern.

\begin{figure}
\includegraphics[width=\columnwidth]{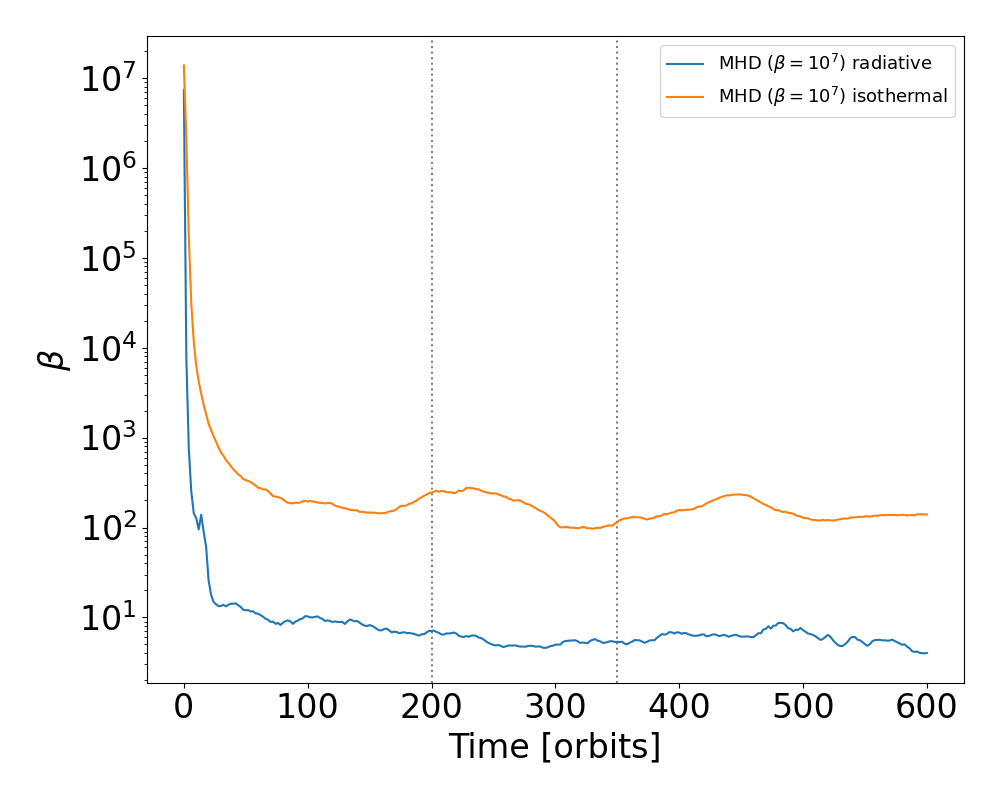}
\caption{Global integrated $\beta$ in two different cases with initial $\beta=10^7$: radiative and isothermal. Despite starting from similar points, the radiative case decreases much more (two orders of magnitude) than the isothermal case. The grey dotted lines represent the moments at which the planet is injected and it is fully grown.}\label{fig:magnetic_energy_iso}
\end{figure}

The different gap-opening mechanism in isothermal and radiative cases comes from the actual magnitude of magnetic fields that develop in the disc. In fact, in the radiative case, the global $\beta$ parameter decreases down to $\sim 10$, while in the isothermal run, $\beta$ never decreases lower than $\sim 10^3$, as shown in Figure \ref{fig:magnetic_energy_iso}. The turbulent motion, caused by the fluctuations of radiative transport, has proven to be more important in radiative HD simulations, with an equivalent $\alpha_R$ that can get almost one order of magnitude above the one shown in the isothermal run, and this causes the magnetic field growth to be much more sustained in the disc. As a consequence, the equivalent $\alpha_M$ is also much stronger and prevents a gap from opening.

\subsection{Accretion of Circumplanetary Material}\label{sec:accretion_CPD}

The different behaviors in the gap-opening process and in the meridional circulation are expected to have consequences on the accretion rates onto the circumplanetary disc and hence onto the planet as well. In order to check that, we followed the amount of mass contained in the Hill sphere around the planet (hereafter defined as $M_{\mathrm{Hill}}$ or \textit{Hill mass}), with the Hill radius defined as
\begin{equation}
    R_H = a \left(\frac{M_p}{3M_\star} \right)^{1/3}
\end{equation}
where $a$ is the semi-major axis of the planet, $M_p$ is the mass of the planet, and $M_\star$ the mass of the star.
We used the Hill-sphere because of the low resolution in the planet's vicinity, which prevents to calculate the accretion rate directly on the planet's surface. However, of course, what material enters the Hill-sphere might not end up on the planet, in fact, it is known that there is a significant outflow from the Hill-sphere based on many previous works \citep{Tanigawa12, Szulagyi14, Ormel15a, Ormel15b, Cimerman17}. 
The value of the Hill mass is of course variable in time as the mass of the planet is also changing during a single simulation and the size of the Hill sphere changes accordingly. For the sake of our analysis, we took into account only the final value, after the disc has reached stability with the fully grown planet.

We studied the evolution of $M_{\mathrm{Hill}}$, comparing the final Hill mass in the HD case with the Hill mass in the four MHD cases (initial $\beta=10^5,10^6, 10^7, 10^8$). Firstly, once again, the MHD case with initial $\beta=10^5$ proves to be different compared to the other MHD cases. In fact, the perturbations at the planet location do not allow the planet to accrete the same amount of mass and a total $M_{\mathrm{Hill}}\simeq2\times 10^{-6} M_\odot \sim 2 \times 10^{-3} M_J$ has been measured. Secondly, the final Hill mass, after the disc has relaxed, is quite different in MHD and HD cases. In fact, a magnetized disc (except of course the case with initial $\beta=10^5$) produces a Hill mass of about $8\times 10^{-6} M_\odot \sim 8 \times 10^{-3} M_J$, where $M_J$ is the Jupiter mass, while the HD case shows a Hill mass that is a factor 4 lower ($2\times 10^{-6} M_\odot \sim 2 \times 10^{-3} M_J$). This is mainly due to the lack of a gap. In fact, the minimum density in the planetary gap is very different in the two cases, as shown in Figure \ref{fig:main_comparison_300}, and even though the circumplanetary density has similar profiles in shape, this offset causes the final value of $M_{\mathrm{Hill}}$ to be higher in the MHD case.

The same analysis can be done when an isothermal treatment is implemented. Since the MHD isothermal case has already shown to give different results compared to its radiative counterpart, we expect the Hill mass to decrease. In fact, the two isothermal cases (MHD and HD) turn out to be indeed extremely similar, not only in the gap opening process but also in the accretion onto the planet, reaching both a Hill mass value slightly below $2\times 10^{-6} M_\odot$. This means that the lower magnetic energy developed in the disc, as shown by Figure \ref{fig:magnetic_energy_iso}, is responsible for the different behaviour of the disc. In general, isothermal runs show a slightly lower Hill mass compared to the radiative HD one, mainly because the disc is slightly warmer around the planet since it cannot radiate energy away (Figure \ref{fig:main_comparison_300_iso}), and this causes the gas to settle less towards the mid-plane.

Lastly, we compared the evolution of $M_{\mathrm{Hill}}$ in the MHD case ($\beta=10^7$) to a variety of HD runs. We have already seen that the isothermal HD run produces much lower values for the Hill mass. At the same time, also the adiabatic setup, despite being able to reproduce the absence of the gap, gives low and comparable values, again because the high temperatures do not allow gas to settle around the planet. On the other hand, radiative and viscous HD setups are much more interesting. In fact, even though the case with $\alpha=10^{-3}$ shows a low mass, at about $4\times 10^{-6}M_\odot$, the $\alpha=10^{-2}$ shows a much higher value, at around $12\times 10^{-6}M_\odot$. This again confirms that effective viscosity is the key factor in the gap-opening process and in planetary accretion as well. Having higher $\alpha$'s not only prevents a gap from fully opening but also fosters higher accretion rates onto the planet. In particular, we can infer that a value for $\alpha$ between $10^{-3}$ and $10^{-2}$ would produce a Hill mass similar to the MHD case.

\begin{figure}
\includegraphics[width=\columnwidth]{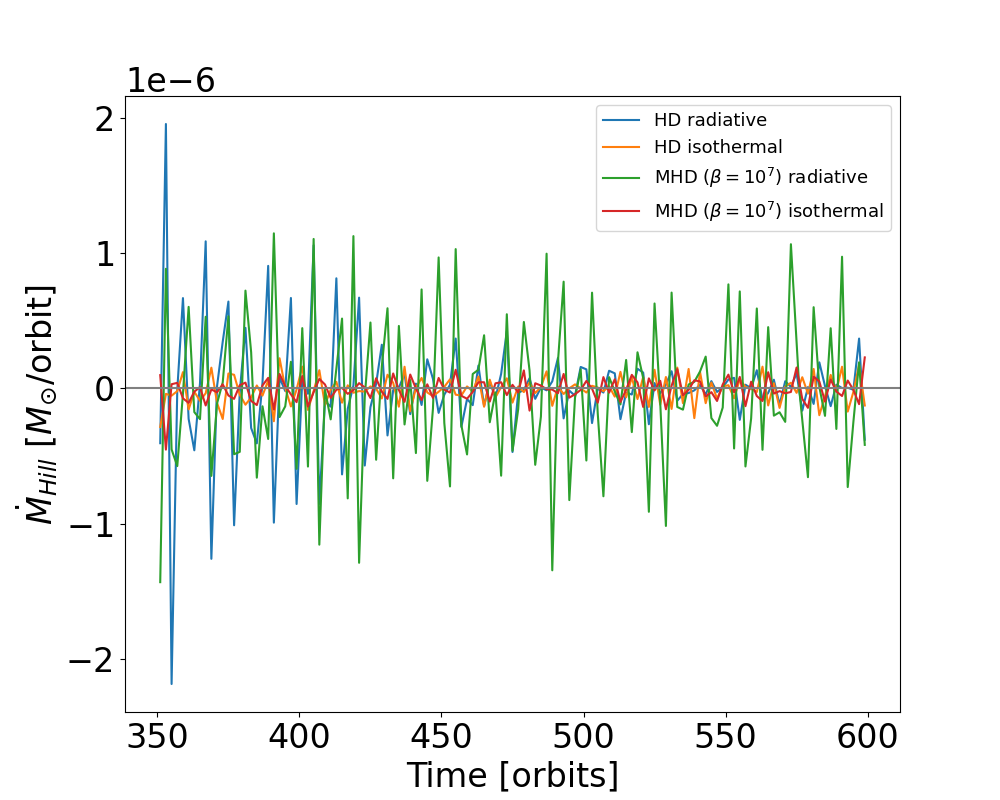}
\caption{Accretion rate onto the Hill sphere after 350 orbits, i.e. when the planet is fully formed. Four different cases are compared: HD isothermal and with radiative transfer, MHD (initial $\beta=10^5$) isothermal and with radiative transfer. Clearly, the isothermal runs are much less variable than the radiative runs, highlighting that the radiative treatment has a large impact on the accretion rate and  variability.}\label{fig:accretion_rates}
\end{figure}

Other than the mass in the Hill sphere, one could measure the mass accretion rate, i.e. the time derivative of the Hill mass itself $\frac{dM_{\mathrm{Hill}}}{dt}$. In general, these values will not be stable in time, nor easily measurable. In fact, shocks and heating/cooling processes around the planet cause the temperature in the Hill sphere to oscillate, and, consequently, also the Hill mass will not have a completely smooth evolution, but will rather show oscillations around a mean value. In fact, when the Hill sphere is colder, more mass can be compressed in it, while when it is warmer, the gas expands and the outer layer is pushed away by thermal pressure \citep{Szulagyi16}, decreasing the overall density of the Hill sphere. This is also why radiative simulations are so important in determining realistic accretion rates.

In particular, in both HD and MHD radiative cases, the accretion rate values initially span from $-10^{-6} \frac{M_\odot}{\mathrm{orbit}}$ to $+10^{-6} \frac{M_\odot}{\mathrm{orbit}}$ with an average value around $+10^{-8} \frac{M_\odot}{\mathrm{orbit}}$ (Figure \ref{fig:accretion_rates}). For comparison, this would translate into an accretion rate of $10^{-6}\frac{M_J}{\mathrm{year}}$, in agreement with the assumed formation timescale of Jupiter. Towards the end of the simulation, after 600 orbits, the oscillation amplitude of the HD cases decreases by about one order of magnitude, reaching the same amplitude values we observed in isothermal cases, both HD and MHD, as shown again in Figure \ref{fig:accretion_rates}. This proves that the combined effect of magnetic fields and radiative transfer makes the meridional circulation, hence accretion, more turbulent and less smooth than all the other combinations.

\subsection{High-resolution Comparisons}\label{sec:high-resolution}

\begin{figure}
\includegraphics[width=\columnwidth]{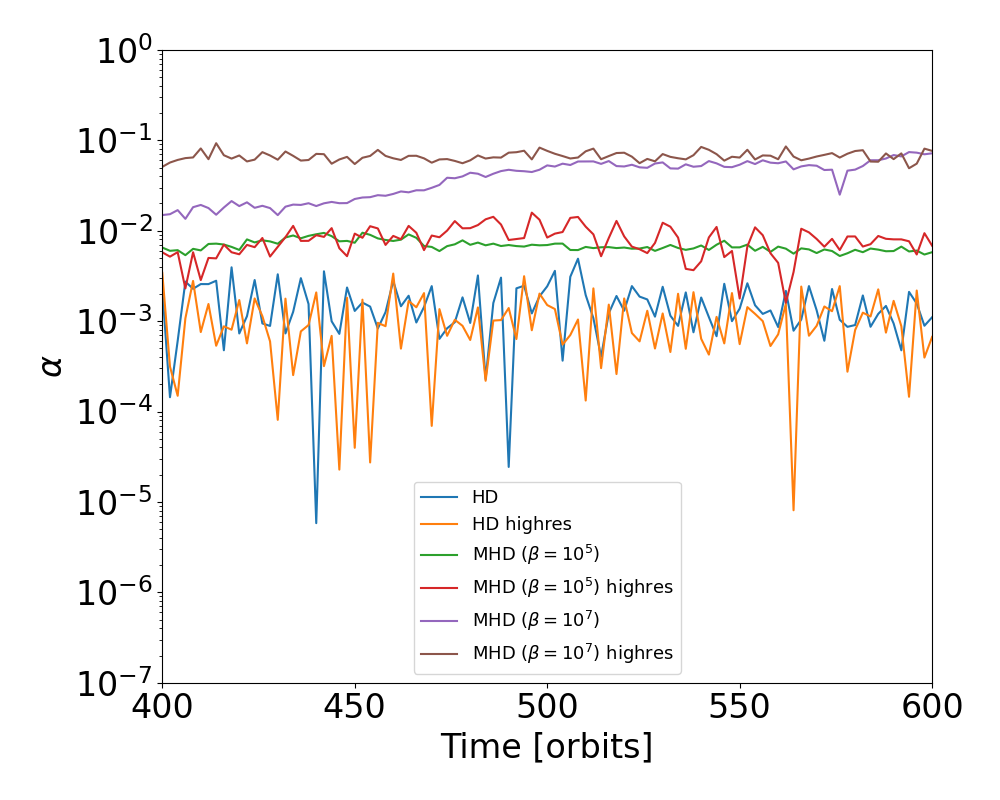}
\caption{Equivalent viscosity in the HD and MHD cases ($\beta=10^5,10^7$) in two different resolutions, from 400 and 600 orbits, after the planet is fully formed.}\label{fig:alpha_highres}
\end{figure}

In order to verify the effect of higher resolutions, we also run some selected configurations in a 420x44x1440 setup (8 times the number of cells as in the low-resolution case). In particular, three different radiative cases were tested. One HD case and two MHD cases with $\beta=10^5,10^7$. Again, an HD case was run for about 50 orbits, and the three setups were initiated from there, obviously adding magnetic fields when necessary. 
The first, and not surprising result, is that the two HD simulation results (low- and high-resolution) are almost identical. In practice, increasing the resolution does not produce relevant effects on the hydrodynamics of the disc. Even the effective viscosity in the disc turns out to be very similar, as shown in Figure \ref{fig:alpha_highres}, with values oscillating around $10^{-3}$ in both resolutions after the planet is formed.

Comparing various resolutions is especially important in MHD since  magnetic effects are known to be sensitive to resolution \citep{Fromang06}. Starting from the equivalent viscosity, we found that in the case with initial $\beta=10^5$, the value for $\alpha$ was roughly the same in the two resolutions (Figure \ref{fig:alpha_highres}) with some more oscillations in the high-resolution run. The latter happens because the residual magnetic fields are the main source of viscosity, in this case. In fact, even though residuals were found not to be significant in the low-resolution runs, in high-resolution we start to resolve the magnetic instabilities, such as the MRI. Consequently, that causes the residual $\alpha$ to be much more important compared to the total viscosity value. This means, in particular, that one should start to see the heating effect of viscosity, which was not important in lower resolution. In the case with initial $\beta=10^7$ though, this is even more relevant. In fact, in this case, the total viscosity was found to be initially lower in the high-resolution run and to reach roughly the same value at the end of the simulation. Given that the total magnetic $\beta$ has found to be even lower than $\sim 10$ at the end of the simulation despite the higher temperature (and thermal pressure), we already have a hint that MHD-driven instabilities could play a major role in this case.

\begin{figure}
\includegraphics[width=\columnwidth]{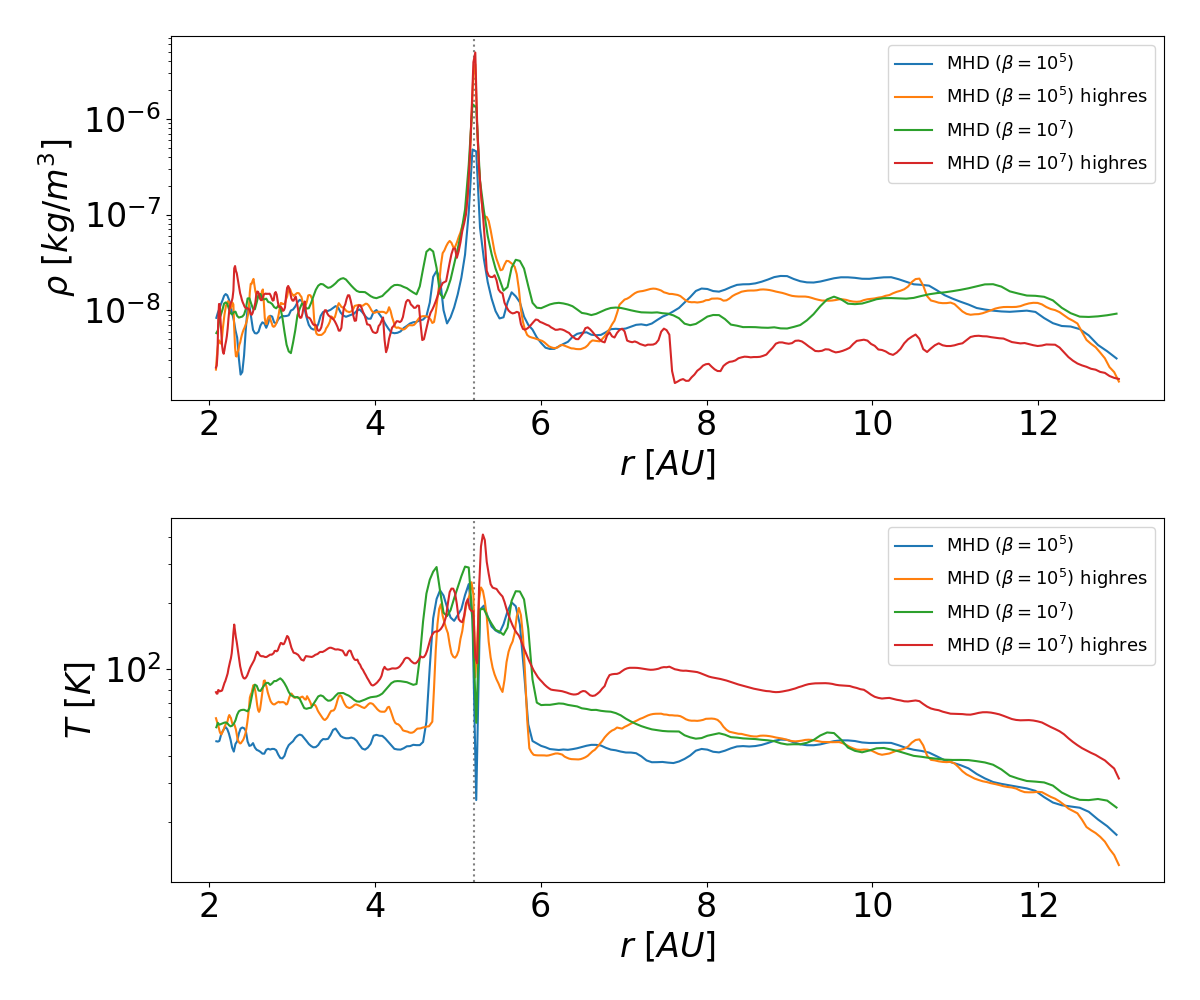}
\caption{Density and temperature profiles in the HD and MHD cases (initial $\beta=10^5,10^7$) in two different resolutions. The grey dotted line represents the planet's location.}\label{fig:main_highres}
\end{figure}

In fact, as shown in Figure \ref{fig:main_highres}, the density and temperature profiles in the $\beta=10^5$ cases are similar enough, with some differences due to a slight overheating in the high-resolution run. On the other hand, heating is much more prominent in the case with initial $\beta=10^7$ because of the residual viscosity, causing the disc to be much hotter, thus the density profile also to be different. This highlights that going to high resolutions is a powerful tool to resolve and detect magnetic instabilities that can modify the outcome of the model in certain circumstances, at least in an ideal-MHD scenario. It is known, that more accurate ionization models predict a low coupling between the gas and the magnetic fields, that damps any MHD-driven instability in the so-called \textit{dead zone}, where planets are forming \citep{Gammie96}. In this more realistic case, increasing the resolution would not resolve any instabilities better and it is expected to give roughly the same results.

\section{Discussion}\label{sec:discussion}

The first study about CPD characteristics and structure with Magneto-Hydrodynamic simulations was presented by \cite{Gressel13}. Their work included Ohmic resistivity (i.e. non-ideal MHD), which leads to turbulent surface layers of the CPD and even a dead zone close to the mid-plane. They  started the simulations with relatively low-intensity vertical magnetic fields (3.6 $mG$, i.e. $\beta \sim 3 \times 10^5$). They found that the accretion flow into the Hill sphere was again dominated by high-latitude inflow, while the gas flowed outwards in the CPD mid-plane. A cavity was opened by magnetic fields between the planet and the inner CPD, similarly to the PPD case between the star and the inner disc edge. In this cavity, the ionization rate was larger, which led to stochastic accretion with a high level of variability. During one instance,  even an intermittent jet was forming because of the magnetic fields. The CPD itself was very light, therefore easier to ionize, while the accretion rate reached a steady value that possibly allows rapid growth from a Saturn- to a Jupiter-mass planet. The low-intensity magnetic field implemented in the model is the one that is generated by the low-ionized PPD, and it is different compared to the magnetic field generated by, for example, the forming planet or the proto-star, even by orders of magnitude.
In agreement with that and with e.g. \cite{Zhu14}, we observed a clear gap opening in our MHD isothermal runs. In fact, as explained in Section \ref{sec:thermodynamics_comparison}, in the isothermal disc case, turbulence in the velocity field is not strong enough to sustain strong growth of the magnetic field's magnitude. This implies that the effective viscosity generated by the Maxwell stress tensor never gets large enough to prevent the gap to open. On the other hand, when we switch on a proper radiative transfer prescription, the effective viscosity generated in the disc is high enough so that a gap is actually prevented to open.

Thermodynamics is not the only factor to affect magnetic field growth. In our case, ideal MHD has been implemented and the absence of a magnetic resistivity is enhancing the MHD effects in the disc and making them dominant in some of its regions, especially at higher latitudes. Non-ideal effects, including Ohmic Resistivity, Ambipolar Diffusion, and the Hall effect, would slow down the magnetic energy growth, possibly preventing the $\beta$ parameter to get to such low values, and, consequently, the effective viscosity to be so large. In order to implement these effects, the ionization fraction of the gas needs to be consistently calculated, as has been shown in previous works.
For example, \cite{Fujii14} studied whether the MRI could develop in CPDs around forming giant planets since it can in principle drive accretion. They found that the vast majority of the volume of the CPD does not meet the criteria for MRI to develop, except for a thin layer at the disc surface. If non-ideal (Ohmic resistivity, ambipolar diffusion, and Hall drift) effects were also considered, the gap region turned out to be very MRI-unstable, while the CPD region did not have high enough magnetic fields that can drive accretion \citep{Keith15}.
Moreover, Ambipolar Diffusion has proven to be a very large effect on the gap-opening process, as shown by \cite{Aoyama23}. In fact, in that case, non-ideal MHD, combined with locally isothermal thermodynamics, has been shown to cause a much deeper gap, highlighting a very different behavior compared to the results of this work, where ideal MHD is implemented together with radiative transfer. This suggests that the combined effect of non-ideal MHD and radiative transfer could produce peculiar results that should be studied in future works.

\begin{figure}
\includegraphics[width=\columnwidth]{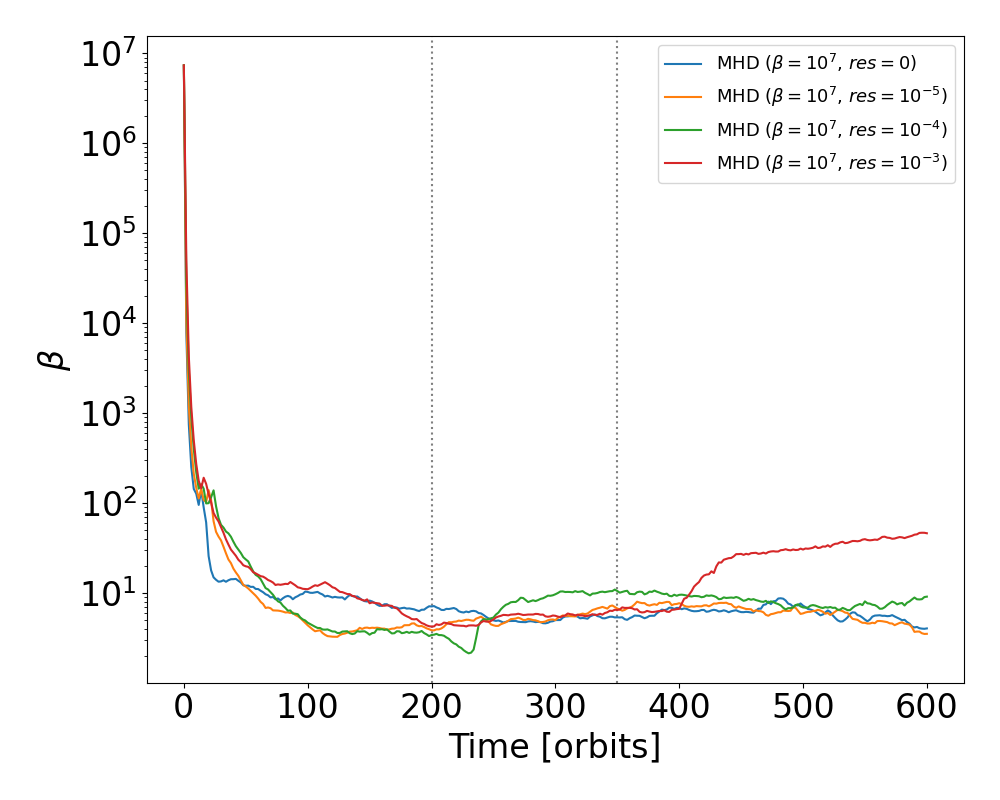}
\caption{Globally integrated $\beta$ (see Equation \ref{eq:beta}) in four different cases, all with initial $\beta=10^7$. The first case is without any buffer zone, then the other three cases have an inner buffer zone with magnetic resistivity $\eta=10^{-5}, 10^{-4}, 10^{-3}$ in code units in the inner ring.} \label{fig:magnetic_energy_res}
\end{figure}

Even without a proper ionization model, Ohmic Resistivity can be used to artificially create a buffer zone in the inner region of the disc, where it should be dominant so that the strong magnetic field growth is damped and slowed down in the region where it is actually most efficient. This way, a smoother evolution of the disc is expected. In particular, we produced other three setups, with initial $\beta=10^7$, but with an artificial buffer zone. The magnetic resistivity has been set to be 
\begin{equation}
\eta = \eta_0\times
\begin{cases}
 \frac{3.12\,AU - r}{1.04\,AU}\;\;\;\;&2.08 \,AU\le r<3.12 \,AU\\
0\;\;\;\;&r\ge3.12 \,AU
\end{cases}
\end{equation}
so that it is equal to $\eta_0$ at the inner ring and to $0$ when the orbital radius reaches $3.12$ AU, i.e. $0.6$ times Jupiter's semi-major axis. Three different values of $\eta_0$ have been chosen, i.e. $\eta_0 = 10^{-5}, 10^{-4}, 10^{-3}$ in code units, chosen so that they have a strong enough effect, but at the same time, they are not too strong so that the CFL condition decreases too much and causes the simulations to slow down significantly. 

\begin{figure}
\includegraphics[width=\columnwidth]{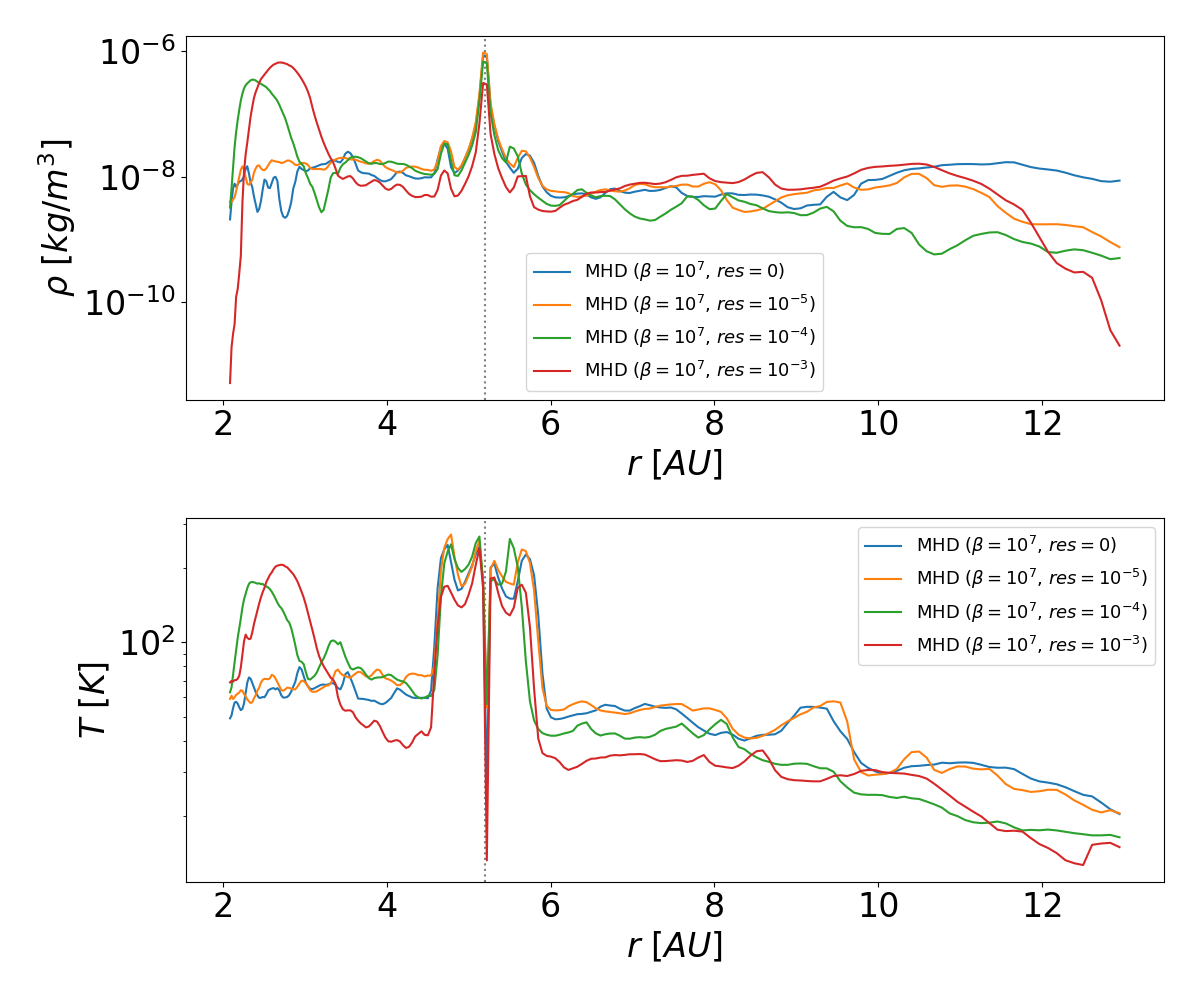}
\caption{Final density and temperature profiles in four different cases, all with initial $\beta=10^7$. The first case is without any buffer zone, then the other three cases have an inner buffer zone with magnetic resistivity $\eta=10^{-5}, 10^{-4}, 10^{-3}$ in code units in the inner ring.} \label{fig:profiles_res}
\end{figure}

As shown in Figure \ref{fig:magnetic_energy_res}, the magnetic energy evolution reaches roughly the same limit (except the case with $\eta_0=10^{-3}$ that proved to be too drastic), but the evolution is smoother and slower. In fact, the first rapid evolution phase shown in Figure \ref{fig:magnetic_energy} is avoided and the disc can evolve while better keeping the equilibrium. Even in this case, the gap does not open, as shown in Figure \ref{fig:profiles_res}, and this was actually expected, given that the integrated $\beta$ reaches the same final value in all the cases. Nevertheless, the final temperature and density profiles are very similar to previous cases, especially around the planet's location. The only significant difference is exactly in the inner disc region, i.e. the buffer zone. Resistivity is heating up the disc and concentrating the density in the mid-plane, while not affecting the disc beyond $r=3.12\,AU$. This buffer-zone strategy has then proven to be useful when even higher resolutions will be explored, and instabilities much better resolved \citep{Fromang06}.

\begin{figure}
\includegraphics[width=\columnwidth]{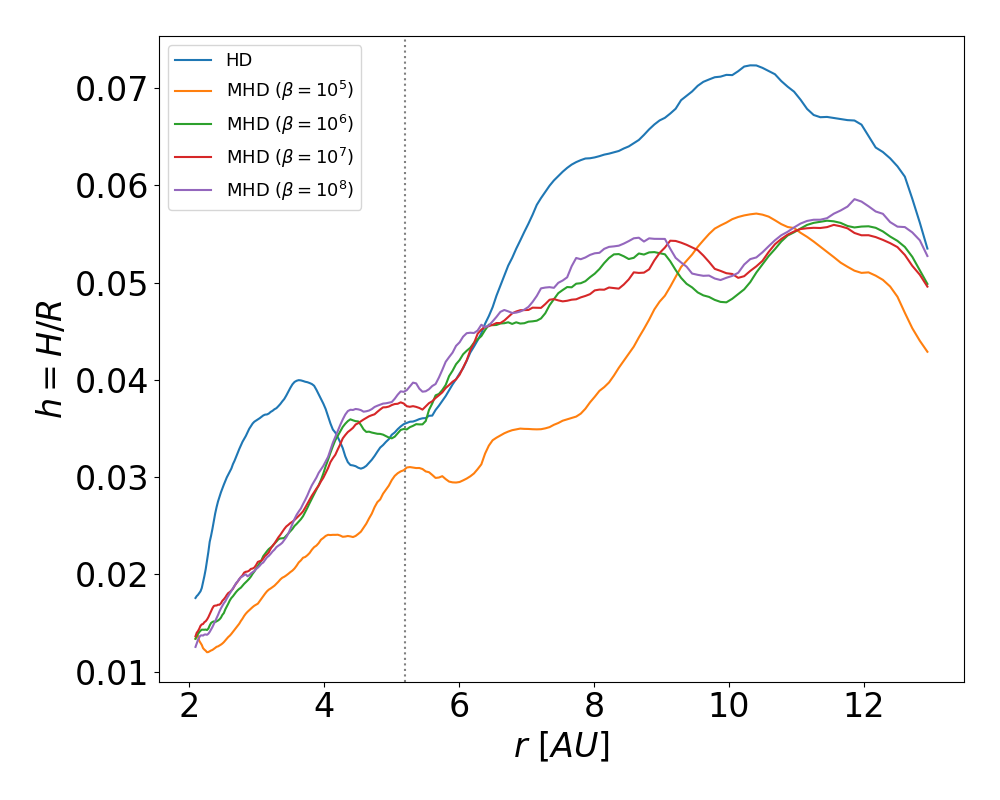}
\caption{Aspect ratio in the disc at the end of the simulation in five different cases: the HD run and the four MHD runs (initial $\beta = 10^5, 10^6, 10^7, 10^8$).} \label{fig:aspect_ratio}
\end{figure}

Furthermore, the chosen resolution can also have an effect on the results. In fact, our low-resolution runs have 22 cells in the vertical direction, with an opening angle of 0.2 radians (only one hemisphere included). The aspect ratio $h$ is initialized at $0.05$ and keeps roughly the same order of value. In fact, as shown in Figure \ref{fig:aspect_ratio}, at the end of the simulation the aspect ratio decreases in the inner disc, then stays at about $0.04$ around the planet, and grows to 0.07 in the outer disc, in the HD setup. This means on average about $5$ cells per scale height in the low-resolution runs, while in the high-resolution runs, we have about 10/11 cells per scale height. This is exactly at the threshold beyond which we should expect to resolve MRI \citep{Gressel13}. In fact, as explained in Section \ref{sec:high-resolution}, in these high-resolution runs, the contribution of the residual Maxwell stress tensor is not negligible anymore when calculating the effective viscosity. We have already seen that the high-resolution runs present more heating due to the residual viscosity, hence different density profiles. We can then expect that increasing the resolution even further would allow MRI to be fully resolved and probably give a stronger contribution to the viscosity in the disc during its evolution, at least in ideal MHD runs. On the other hand, another consequence of non-ideal effects would be causing lower ionization rates in the disc, and then the existence of a dead zone in which MRI should not operate efficiently, regardless of the resolution.

Increasing the resolution would be useful nevertheless, especially around the planet. So far we could only study the accretion rates and the mass in the Hill sphere, which is the most reasonable volume our resolution allows us to investigate. In fact, in the low-resolution runs only 8 cells are included in the Hill sphere in the radial direction, while we reach the number of 16 in the high-resolution cases. This means we do not have enough resolution to really distinguish and resolve the structure of a forming circumplanetary disc, and consequently, we cannot argue about the MHD effects on its evolution and calculate the proper planetary accretion rates, until the planetary radius is resolved. Of course, increasing the resolution of the whole setup would cause the simulation to slow down massively, both because of the higher number of cells and the shorter time steps needed. Two possible solutions could be implemented in this case. The first one would be using a nested grid structure (e.g.  \citealt{Szulagyi16}.). On one hand, this method is convenient because it maintains lower resolution in most of the disc, while it focuses the simulation in the region around the planet. On the other hand, spiral wakes in the disc are fundamental in order to understanding and studying meridional circulation and accretion rates. Therefore an adaptive-mesh-refinement method (AMR), which is already included in PLUTO \citep{Mignone12a} but not compatible yet with the radiative transfer module, would be ideal in this case since the grid would be built in order to increase resolution where the density is higher.

To conclude with, the meridional circulation does not affect gas alone, but also dust and solid particles follow such a pattern during the formation of a giant planet. HD simulations showed in fact that a gap is opened in the dust density distribution as well, at least for the smaller grain sizes \citep{Dipierro17, Binkert21, Szulagyi22}, with the gas circulation preventing dust particles to settle in the mid-planet and creating a thicker solid disc, compared to discs with no planet in them. As the gas is delivered to the CPD and accreted onto the forming planet, also dust particles follow a similar pattern. On average, the dust flow has a dust-to-gas ratio of about 1 \% ($10^{-6}-10^{-8} M_{\rm{jup}} yr^{-1}$ for gas, compared to $10^{-8}-10^{-10} M_{\rm{jup}} yr^{-1}$ for dust), but solid particles can get trapped easier and result in a solid-enriched circumplanetary disc \citep{Szulagyi22}. Even in this case, the implementation of ideal and non-ideal MHD would give a more realistic representation of the circulation and of enrichment of CPDs.

\section{Conclusion}\label{sec:conclusion}

In this work, we investigated the meridional circulation \citep{Szulagyi14, Szulagyi21}, the gap opening process, and the Hill-sphere accretion rates in radiative and magnetized protoplanetary discs using the PLUTO code \citep{Mignone07}. We set up a protoplanetary disc with an initial aspect ratio of $h=0.05$ (evolving then to values between $0.03$ and $0.04$) and a Jupiter-mass planet forming at 5.2 AU. MHD has been implemented in its ideal form, with no resistivity or diffusion. Radiative transfer has been solved by introducing the radiation energy variable, and its evolution has been implicitly solved via a supplementary module of PLUTO \citep{Flock13}. Different simulation setups have been run. In particular, we chose four different values for the initial $\beta$ parameter. Furthermore, in some of the runs, we implemented viscosity, via a proper $\alpha$ parameter \citep{Shakura73}, and we also switched off radiative transfer in order to compare the results with isothermal treatment.

\begin{itemize}

\item The first preliminary result we obtained is that the MHD simulations with initial $\beta=10^6, 10^7, 10^8$ all produced a very similar disc evolution pattern, reaching the same final magnetic strength values for the globally integrated $\beta$ ($\sim 10$), and similar final density and temperature profiles of the disc. In the case with initial $\beta=10^5$, the evolution of the magnetic field is faster at the beginning of the simulation, causing perturbations to propagate from the inner disc up to the planet location, which affects the density and temperature in this area, and, consequently, the structure of the meridional circulation and of the gas accretion onto the Hill-sphere.

\item In both the HD and MHD runs, we retrieved the pattern of \textit{meridional circulation} \citep{Szulagyi14, Szulagyi22}. In the PPD the planetary spiral wakes stir the disc material vertically, while in the planetary gap region, the spirals bring in material to bridge over the gap and deposit material onto the circumplanetary region from the vertical direction. The gas which could not be accreted by the planet right away will flow out from the CPD and goes back into the circumstellar disc.  On the other hand, the mid-plane velocity pattern is very different in HD and MHD cases. In fact, without magnetic fields, it is possible to recognize rings of sub- and super-Keplerian gas between where dust accumulation should be more likely \citep{Kato10}. The azimuthal structure is not so well-defined in the MHD runs, thus making the concentration of solid particles in rings significantly less likely.

\item We calculated the effective $\alpha$ parameter with the help of the Reynolds stress tensor in all simulations, in order to quantify the angular momentum loss mechanisms in the disc. This way, we were able to quantify the effective viscosity due to magnetic processes vs. the effects of the planet's spiral wakes. On one hand, when the gas passes through the spiral shock front, it loses angular momentum. Furthermore, the stirring effect of the meridional circulation also adds to the angular momentum transport in the disc. We found that, in the HD simulations where the magnetic fields do not play any role, the spiral wakes of planets contribute to a high $\alpha\sim 10^{-3}$, highlighting this could be one of the most important momentum loss mechanisms in a realistic PPD when a massive enough planet present. In the MHD simulations, $\alpha$ starts at about $\sim 10^{-2.5}$ before the planet is inserted (hence there are no spiral wakes in the disc) and grows up to $\sim 10^{-1.5}$ at the end of the simulation with the planet fully formed. This means that expectedly, the contribution of magnetic fields is  creating significant turbulence in the disc in the ideal MHD limit, next to the effective viscosity contribution of the planetary spiral wakes.

\item The gap opening process has been found to be extremely different in radiative HD and radiative MHD cases. In fact, while a clear gap was opened in the HD runs, the MHD simulations showed a total absence of it. This can be related to the much higher effective viscosity in the disc in the MHD case. Analytical estimates show that such high viscosity values counteract and win over the effect of planetary torque, stopping the gap opening. This effective viscosity is given by the mean-field stress tensor. This means that while it provides an efficient mechanism to transport mass and momentum, it does not trigger significant viscous heating in the disc.

\item The effective viscosity is significantly lower in isothermal runs than in radiative simulations. In fact, the larger turbulence caused by radiative processes triggers a very fast magnetic field growth, especially from the inner disc and from higher latitudes, where magnetic fields are dominant, given the low density. The lower effective viscosity in the isothermal runs allows the gap to clearly open. This highlights that the inclusion of heating and cooling effects is very important to correctly simulate disc evolution and planet formation.

\item The Hill-sphere mass and accretion rate were also compared in the different simulations. Given the low densities caused by the gap opening process, the radiative HD runs showed a much lower \textit{Hill mass}, about a factor 4 lower than MHD cases, but the two setups revealed comparable Hill masses in isothermal runs. This means again that magnetic fields become dominant in radiative simulations and will cause a different evolution of the disc. At the same time, viscous HD runs showed to be able to produce higher Hill masses, similar to or even larger than the radiative MHD cases. In both the radiative HD and radiative MHD runs, the Hill-sphere accretion rate values oscillate between $-10^{-6} \frac{M_\odot}{\mathrm{orbit}}$ to $+10^{-6} \frac{M_\odot}{\mathrm{orbit}}$ with an average value around $10^{-6}\frac{M_J}{\mathrm{year}}$. This is because due to the heating and cooling effects the Hill-sphere expands and contracts, resulting in the oscillation between the accretion rate sign. 

\item The High-resolution runs started to resolve MRI, providing a higher residual viscosity, especially in the case with initial $\beta=10^7$. This causes  the disc to develop more heating compared to the low-resolution runs, even though the total viscosity is found to be lower. This underlines the need to use high resolutions that can resolve all physical mechanisms, including the fundamental instabilities developing in ideal MHD.

\end{itemize}

\section*{Acknowledgements}

The authors sincerely thank Ruobing Dong for his thoughtful and very useful referee report. M.C.'s work has been carried out within the framework of the National Centre of Competence in Research PlanetS supported by the Swiss National Science Foundation under grants 51NF40\_182901 and 51NF40\_205606, and it was funded by the University of Zurich through the Candoc Forschungskredit grant K-76105-04-01. J.Sz. thanks for the funding from the European Research Council (ERC) starting Grant (No. 948467). M.F. acknowledges funding from the European Research Council (ERC) under the European Union’s Horizon 2020 research and innovation program (grant agreement No. 757957).
Computations partially have been done on the "Piz Daint" and on the "Eiger" machines hosted at the Swiss National Computational Centre (CSCS).
\section*{Data Availability}

The data underlying this article will be shared on reasonable request to the corresponding author.



\bibliographystyle{mnras}
\bibliography{bibliography} 




\appendix

\section{MHD case: $\beta=10^5$}\label{sec:beta_1e5}

\begin{figure}
\includegraphics[width=\columnwidth]{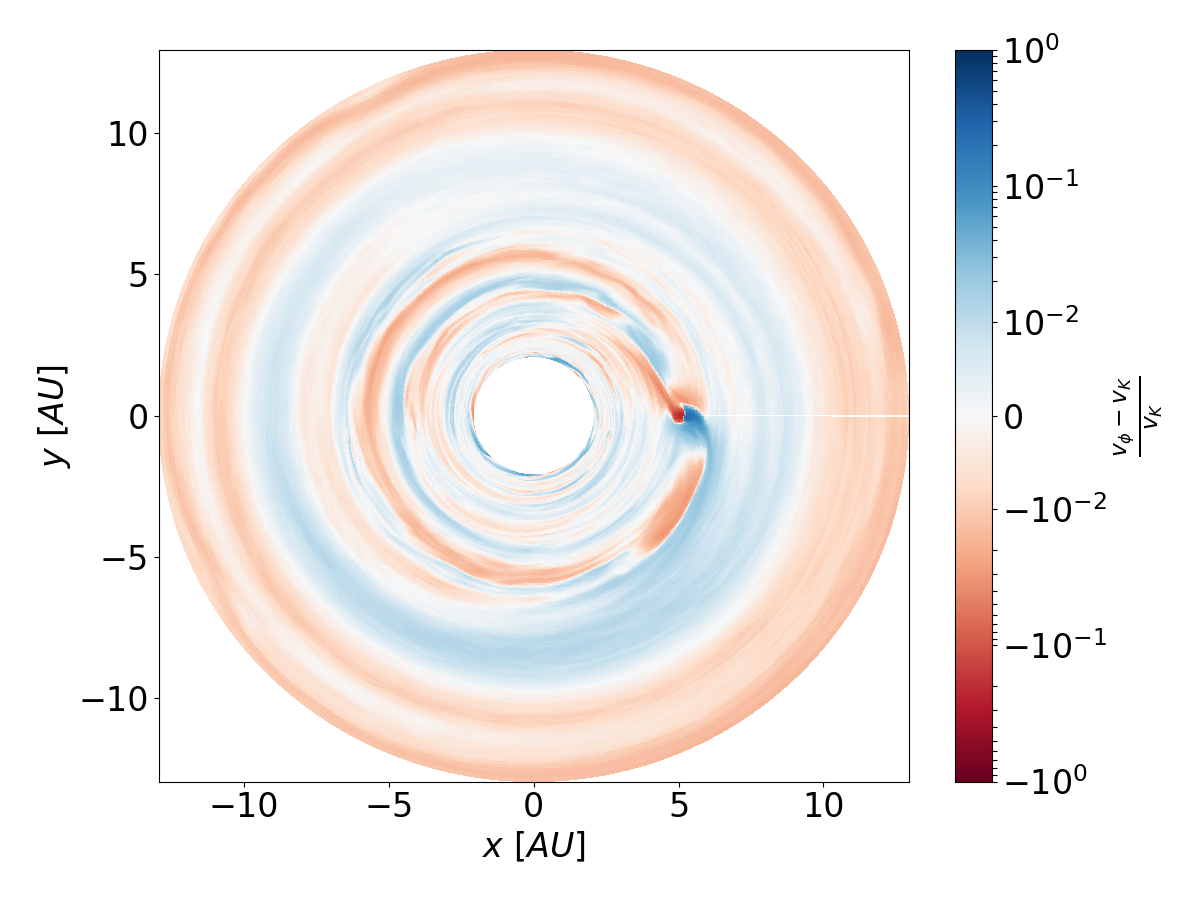}
\caption{The plot shows the difference between the local Keplerian velocity and the actual azimuthal velocity, normalized to the Keplerian velocity itself, in the MHD case with initial $\beta=10^5$. Red areas represent sub-Keplerian motion, while blue ones represent the super-Keplerian flow. }\label{fig:vel_diff_MHD_1e5}
\end{figure}

\begin{figure*}
\includegraphics[width=0.7\textwidth]{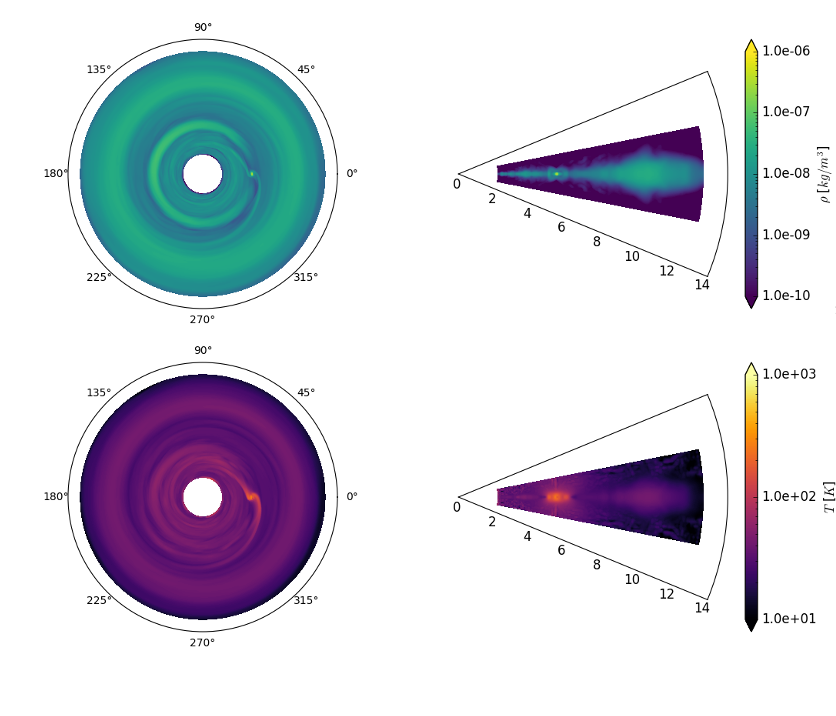}
\caption{Section of the disc at the end of the MHD run with initial $\beta=10^5$. On the left mid-plane sections are represented, while on the right vertical sections at the planet's location are. The top row shows density and the bottom row shows temperature.}\label{fig:main_MHD_1e5}
\end{figure*}

\begin{figure*}
\includegraphics[width=0.95\textwidth]{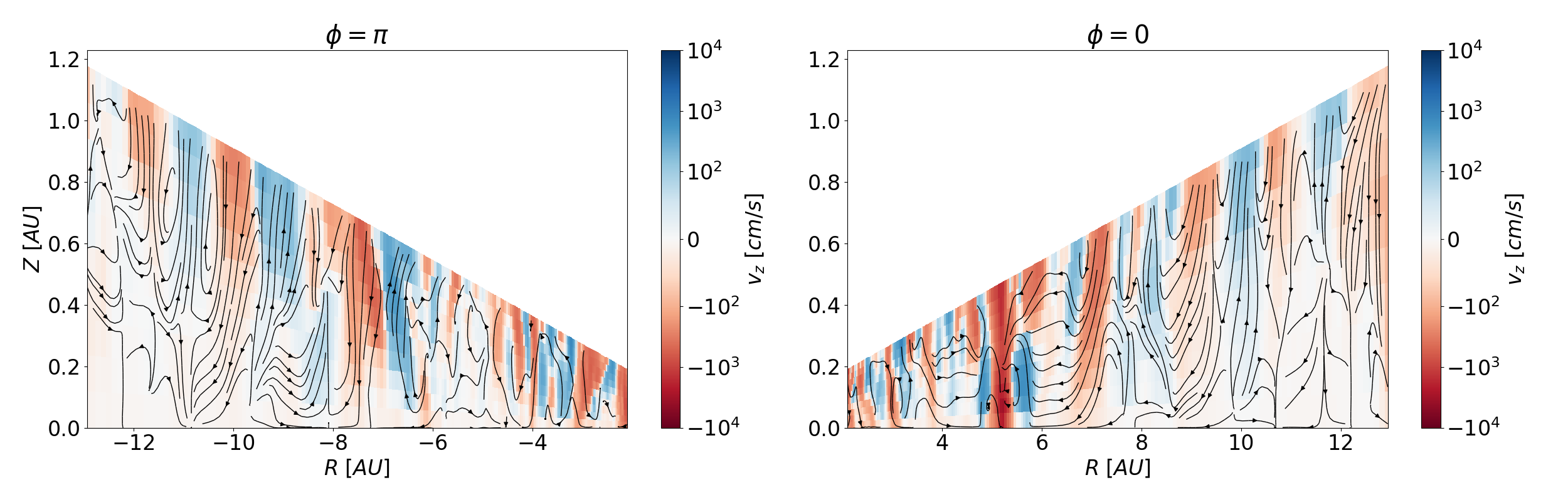}
\caption{Two vertical sections of the protoplanetary disc after 300 orbits, with no magnetic fields, when the planet has already fully formed and the disc has already fully relaxed. The sections are taken at $\phi=0$ and $\phi=\pi$ with respect to the position of the planet, meaning that one section corresponds to the planet's location, and the other is the opposite. The arrows represent the direction of gas velocity, while the color scale represents the value of the vertical velocity component, whether it is positive or negative. Only the upper hemisphere is shown.}\label{fig:MHD_1e5_meridional_circulation}
\end{figure*}

In Section \ref{sec:results}, we chose the MHD case with initial $\beta=10^7$ as the reference case to analyze and compare to the HD runs. This is because lower $\beta$'s caused the magnetic evolution to be quite fast and to produce perturbations that developed from the inner disc up to the planet's orbit. Even though these perturbations may affect the gap-opening process and the measure of the accretion rate, they appear in the inner disc even when resistivity is implemented in order to create a buffer zone (see Section \ref{sec:discussion}). This suggests that their origin is physical rather than numerical, hence it is worth it to mention even the cases with lower initial $\beta$'s.

First of all, Figure \ref{fig:vel_diff_MHD_1e5} shows the sub- and super-Keplerian areas  in the disc. In this case, the ring structure is certainly not as well-defined as in the HD run, but at the same time, the boundaries are much more clear than in the MHD case with initial $\beta=10^7$. This means that the perturbation propagating from the inner disc is a dominant effect in shaping the disc structure and it mostly determines whether the gas is sub- or super-Keplerian, except in the close vicinity of the planet.

This is even more clear when the gap-opening process is investigated.  Figure \ref{fig:main_MHD_1e5} shows the density and temperature maps (both horizontal in the mid-plane and vertical at the planet location) of the disc after the simulation ended. Even in this case, the result is not similar to the HD case nor to the other MHD cases. Even though the temperature profile is similar to other MHD cases, the gap is still not opening. On the contrary, more mass is concentrated along the planetary orbit. This means again that the effect of the inner perturbation is dominant in how the disc structure is shaped. As a result, the Hill mass is in this case lower than other magnetized cases, almost identical to the HD run (see Figure \ref{fig:main_comparison_300}).

To end with, Figure \ref{fig:MHD_1e5_meridional_circulation} shows the velocity direction in two vertical sections of the disc, similarly to Figure \ref{fig:HD_meridional_circulation} and \ref{fig:MHD_meridional_circulation}, and the magnitude of the vertical component of the velocity itself. Even though the meridional circulation pattern is still visible, the magnitude of the velocity is stronger just around the planet's location. At the same time, the rest of the disc is less affected by turbulence and circulation. Once again, the magnetized case with initial $\beta=10^5$ turns out to have similarities with the HD case. This is expected, because, even though starting with the highest magnitude, the magnetic fields in the $\beta=10^5$ are not as strong as in the other MHD cases at the end of the simulation (see Figure \ref{fig:magnetic_energy}), but at the same time, the equivalent viscosity is stronger (see Figure \ref{fig:alphav_comparison} and \ref{fig:alpha_comparison}). This causes the observed hybrid behavior, with the velocity pattern, temperature profiles, and Hill mass being similar to the HD case, the absence of a gap confirming MHD cases, and density profiles and the sub-/super-Keplerian structure being peculiar and not really comparable to previous cases.

\section{PLUTO vs JUPITER validation}\label{sec:PLUTO_JUPITER}

\begin{figure*}
\includegraphics[width=0.8\textwidth]{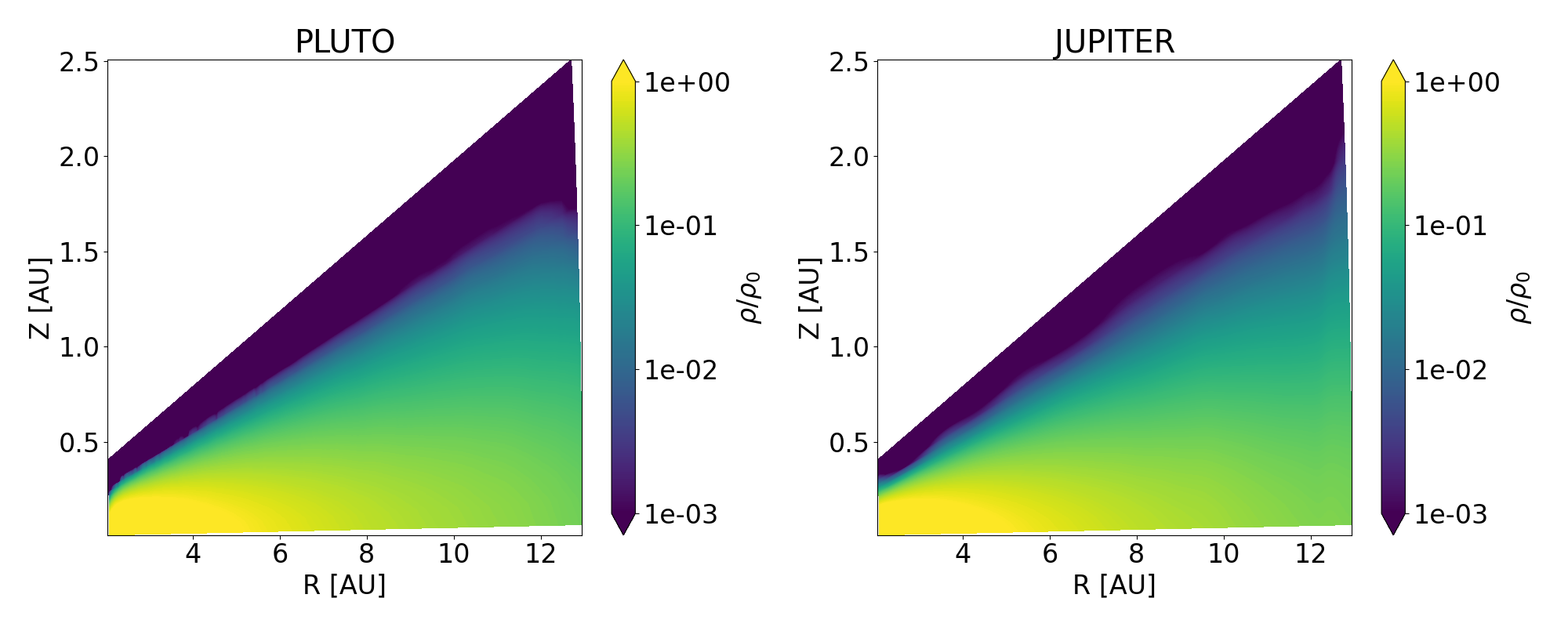}
\caption{Volume density comparison between PLUTO and JUPITER after 20 orbits. The density is normalized to $\rho_0$, i.e. the initial volume density at the planet's location.}\label{fig:PLUTO_JUPITER_density}
\end{figure*}

\begin{figure*}
\includegraphics[width=0.8\textwidth]{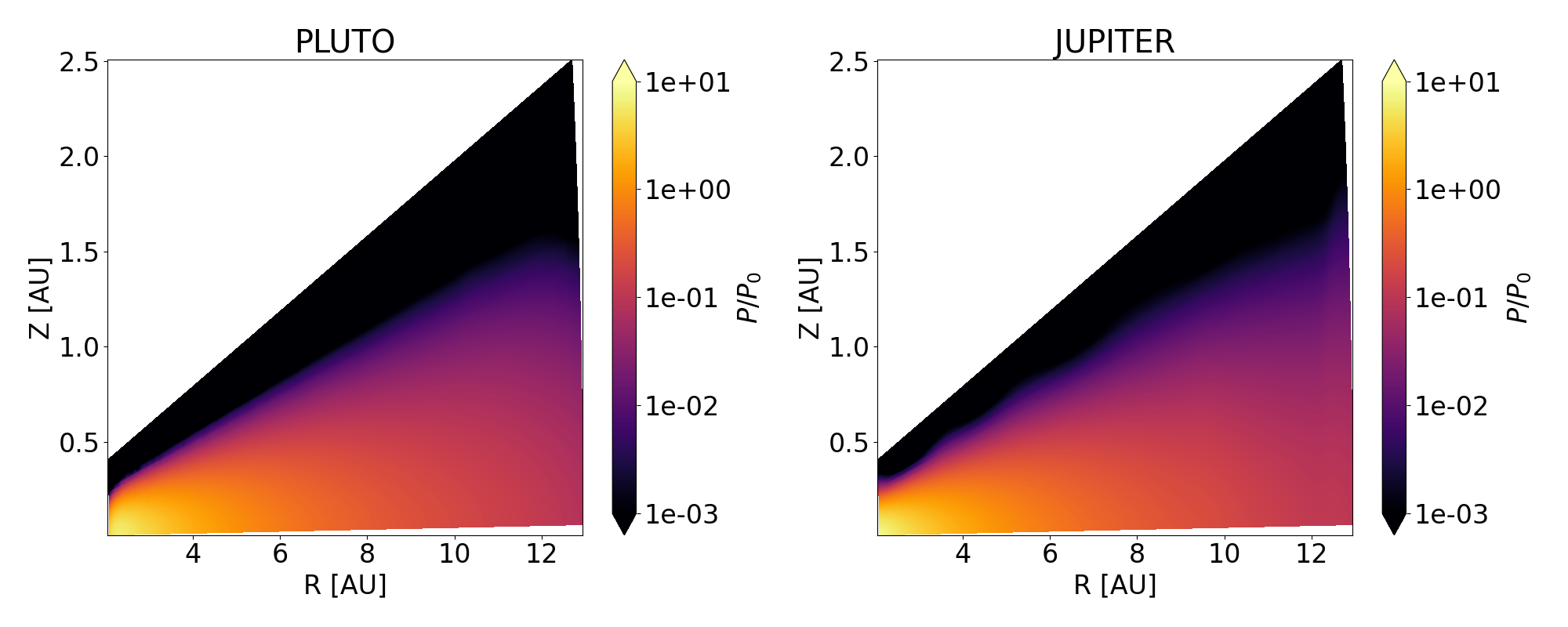}
\caption{Pressure comparison between PLUTO and JUPITER after 20 orbits. The pressure is normalized to $P_0$, i.e. the initial pressure at the planet's location.}\label{fig:PLUTO_JUPITER_pressure}
\end{figure*}

\begin{figure*}
\includegraphics[width=\textwidth]{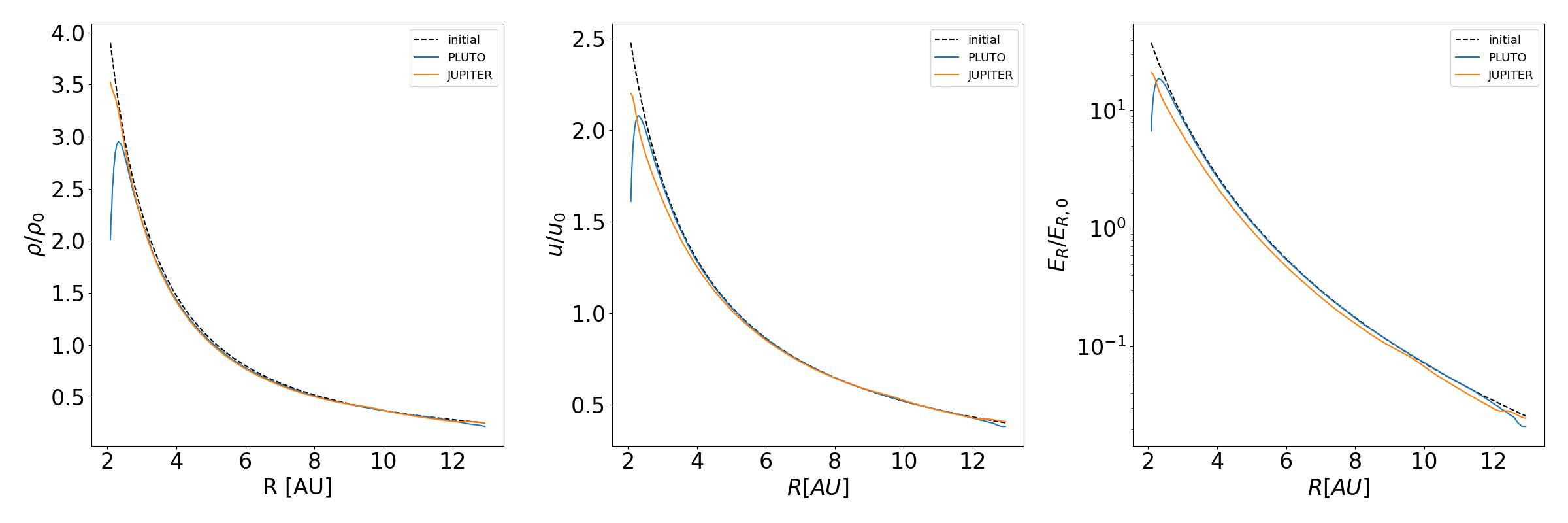}
\caption{Comparison of the mid-plane profiles of the density $\rho$, the internal energy $u$ and the radiation energy $E_R$. The dotted line represents the initial condition.}\label{fig:PLUTO_JUPITER_profiles}
\end{figure*}

As mentioned in Section \ref{sec:numerical_methods}, the radiative transfer solver of PLUTO \citep{Mignone07, Flock13} has been validated by comparing  equivalent runs with the JUPITER code \citep{Szulagyi16} that includes a radiative transfer solver as well.  

In particular, we run the same setup with the same initial conditions and boundary conditions for 20 orbits, without a planet injected in the simulation. Since this assures cylindrical symmetry, then the tests were run in a 2D configuration, where we only considered the $r$ and the $\theta$ directions. We considered 200 cells in the radial direction, spanning from 2.08 AU to 13 AU, while 40 cells were considered in the meridional direction, from $\theta=1.37$ to $\theta=1.77$, i.e. symmetrically around $\pi/2$. The radial cells were logarithmically spaced in PLUTO, while linearly spaced in JUPITER. The same boundary conditions and initial conditions described in Section \ref{sec:methods} were implemented, while to simplify the problem, a constant opacity $\kappa_P = \kappa_R = 10 cm^2 g^{-1}$ was implemented. Some time was invested in order to make sure that the two set-ups were as close as possible, despite the difference between the two codes.

First, we visually compared the 2D density and pressure distribution in the $R\times Z$ space. At the beginning of the simulation, both the density and pressure profiled vary smoothly from the mid-planet to higher latitudes, following the exponential decay described in Section \ref{sec:disc_model}. After 20 orbits, one can see that both the density and pressure profiles reached the same equilibrium configuration. Figure \ref{fig:PLUTO_JUPITER_density} and \ref{fig:PLUTO_JUPITER_pressure} show in fact that with both codes the disc contracts and develops a sharper upper boundary, while maintaining very similar profiles close to the mid-plane.

When analyzing the mid-plane profiles of density, internal energy, and radiation energy, as shown in Figure \ref{fig:PLUTO_JUPITER_profiles}, one can also see that they are in fact very similar away from the boundaries, with differences smaller than a few percents. On the other hand, one can also observe that the two radiative transfer solvers behave slightly differently when dealing with ghost cells close to the boundaries. 
In fact, the two different grid structures influence the optical thickness in each cell and consequently, the cooling efficiency at the boundary. 
Nevertheless, since we are mainly interested in the disc evolution around the planet location at 5.2 AU, we can claim that the behavior of the two codes is equivalent within the error bars, and our radiative method in PLUTO is validated.


\bsp	
\label{lastpage}
\end{document}